\documentclass[a4paper]{jpconf}
\usepackage{graphicx}

\def\Journal#1#2#3#4{{#1}{\bf #2}, #3 (#4)}

\def\EPJC{{Eur. Phys. J.}~{\bf C}}

\def\NIMA{{Nucl. Instrum. Methods}~{\bf A}}
\def\NPA{{Nucl. Phys.}~{\bf A}}
\def\NPB{{Nucl. Phys.}~{\bf B}}
\def\PLB{{Phys. Lett.}~{\bf B}}
\def\PLC{Phys. Repts.\ }
\def\PRL{Phys. Rev. Lett.\ }
\def\PRD{{Phys. Rev.}~{\bf D}}
\def\PRC{{Phys. Rev.}~{\bf C}}
\def\ZPC{{Z. Phys.}~{\bf C}}
\def\ZPA{{Z. Phys.}~{\bf A}}
\newcommand \etaL{ {\it et al.} } 

\newcommand \mt{$\langle m_{T}\rangle$ }
\newcommand \Nch{$N_{ch}$ }
\newcommand \Et{$E_{T}$ }
\newcommand \EN{$E_{T}/N_{ch}$ }
\newcommand \dNch{$dN_{ch}/d\eta$ }
\newcommand \dEt{$dE_{T}/d\eta$ }
\newcommand \Np{$N_{p}$ }
\newcommand \Nps{$N_{p}$}
\newcommand \sqn{$\sqrt{s_{_{NN}}}$ }
\newcommand \sqns{$\sqrt{s_{_{NN}}}$}
\newcommand \1{19.6~GeV}
\newcommand \2{200~GeV}
\newcommand \3{130~GeV}
\newcommand \6{62.4~GeV}
\newcommand \7{17.2~GeV}

\begin{document}

\title{Centrality and \sqn Dependence of the \dEt and \dNch in Heavy Ion Collisions at Mid-rapidity}

\author{A.~Milov for the PHENIX Collaboration\footnote{For the full list of authors and acknowledgments see reference~\cite{ppg19}}}
\address{Department of Physics and Astronomy, Stony Brook University, SUNY, Stony Brook, NY 11794, USA}

\begin{abstract}
The PHENIX experiment at RHIC has measured transverse energy and charged particle multiplicity 
at mid-rapidity in $Au+Au$ collisions at \sqn = 19.6, 130, 62.4 and 200 GeV as a function of centrality.
The presented results are compared to measurements from other RHIC experiments, and experiments at 
lower energies. The \sqn dependence of \dEt and \dNch per pair of participants is 
consistent with logarithmic scaling for the most central events. The centrality dependence of 
\dEt and \dNch is similar at all measured incident energies. At RHIC energies the ratio of transverse 
energy per charged particle was found independent of centrality and growing slowly with \sqns. 
A survey of comparisons between the data and available theoretical models is also presented.
\end{abstract}

\section{Introduction}
The PHENIX experiment at the Relativistic Heavy Ion Collider (RHIC) at Brookhaven National Laboratory was designed to
measure the properties of matter at extremely high temperatures and densities. Under such conditions, there is possibility 
to produce states of matter that have yet to be observed or studied in the laboratory. The best known
of these is the {\em quark-gluon plasma} (QGP), a form of matter where quarks are not confined within individual baryons
but exist as some form of plasma of individual quarks and gluons. It should be emphasized that the exact properties of
this matter are not known and that the characterization of the deconfined state, if such a state is produced, is
an essential part of the RHIC program.

One fundamental element of the study of ultrarelativistic collisions is the characterization of the interaction in terms
of variables such as the energy produced transverse to the beam direction and the number of charged particles. These 
two variables are closely related to the collision geometry and are important in understanding global properties of the 
system during the collision.

This paper describes a systematic study of \dEt and \dNch at mid-rapidity\footnote{Also referred as \Et and \Nch 
in this paper.} by the PHENIX experiment at center-of-mass 
energies \sqns~=~19.6, 130, 62.4 and \2. The centrality dependence of \Et and \Nch is characterized by the number of 
participants, determined with a Glauber model, and is studied as function of the incident energy.
\Et and \Nch results for all four RHIC measurements are included as part of this study.
The data taken at 19.6 GeV is particularly interesting because it allows a close comparison to the lower energies 
of the CERN SPS program. Comparisons are also made to previous experiments at the Brookhaven AGS 
and CERN SPS at center-of-mass energies of 4.8 GeV, 8.7 GeV, and 17.2 GeV. Finally, an extensive set of collision models
describing the \Et and \Nch distributions are compared to the existing data.

\section{PHENIX detector and Analysis\label{sec:dtector}}
PHENIX is one of four experiments located at RHIC~\cite{rhic}. 
The PHENIX detector consists of two central spectrometer arms, 
designated east and west for their location relative to the interaction region,
and two muon spectrometers, similarly called north and south.  
Each central spectrometer arm covers a rapidity range of $|\eta|<0.35$ and subtends $90^{0}$ in azimuth. 
The  muon spectrometers both have full azimuthal coverage with a rapidity ranges of 
$-2.2<\eta<-1.2$ (south) and $1.2<\eta<2.4$ (north). Additional global 
detectors are used as inputs to the trigger and for global event characterization 
such as vertex, time of event and centrality determination.
A detailed description of the PHENIX detector can be found in~\cite{phenix}. 

The PHENIX detector subsystems relevant for the physics analysis published here are:
the Pad Chambers used for the charged particle multiplicity measurement, the 
Electromagnetic Calorimeter used to measure transverse energy, the Beam-Beam Counter 
and the Zero Degree Calorimeter, the two detectors used for triggering and centrality 
determination are described in publications~\cite{pc,emcal,bbc-zdc,zdc}.

The analysis procedures to measure \Et and \Nch are described in details in 
publications~\cite{phenix_milov,phenix_bazik,david_thesis,sasha_thesis}. Some additional
information concerning the analysis at \sqns~=~\1 can be found in~\cite{ppg19}. The preliminary 
results on \dNch at \sqns~=~\6 use the same technique as implemented at other energies. 

For the sake of space we only mention errors relevant to the analyses. Statistical errors are 
small and do not exceed 1\% of the measured value. The systematic errors are summarized in 
Table~\ref{tab:errors}. The errors for the lowest and the highest incident energy are listed, 
whereas at all other energies the are close to these values. The systematic errors 
for both measured values are of two types. The first type affects the centrality 
dependence. It is listed in the Table~\ref{tab:errors} with the range 
(first number corresponds to the most central bin). Errors of the second type contribute 
to the overall scaling of the data. In the figures below the centrality dependent errors are 
shown as a corridor, and the the full systematic error is shown with the 
error bars\footnote{Here and everywhere errors correspond to one standard 
deviation.}. Centrality related errors are common to both measured quantities.

\begin{center}
\begin{table}[h]
\caption{Summary of systematic errors given in \%. When the range is given, 
the first number corresponds to the most central bin and the second to the 
most peripheral bin.
\label{tab:errors}}
\centering
\begin{tabular}{lcccc}
\br
\multicolumn{1}{c}{}&\multicolumn{2}{c}{\dEt}  & \multicolumn{2}{c}{\dNch}\\
\sqn [GeV]          &  19.6    & 200    & 19.6     & 200     \\
\mr
Energy resp.        &  4.7     & 3.9    &          &         \\
Bkg. / noise        &  0.5-3.5 & 0.2-6  & 1.0      & 1.0     \\
Acceptance          &  2.0     & 2.0    & 2.3      & 2.3     \\
In- \& outflow      &  3.0     & 3.0    & 5.7      & 2.9     \\
Occupancy           &          &        & 1.6-0.3  & 3.5-0.1 \\
\hline			       					
Centrality          & 2.0      & 0.5    & \multicolumn{2}{c}{same} \\
\Np                 & 2.9-6.7  & 2.8-15.& \multicolumn{2}{c}{same} \\
Trigger             & 0.4-8.8  & 0.3-16.& \multicolumn{2}{c}{same} \\
\br
\end{tabular}
\end{table}
\end{center}

\section{Results \label{sec:result}}
\subsection{PHENIX results}
\begin{figure}[h]
\includegraphics[width=0.49\linewidth]{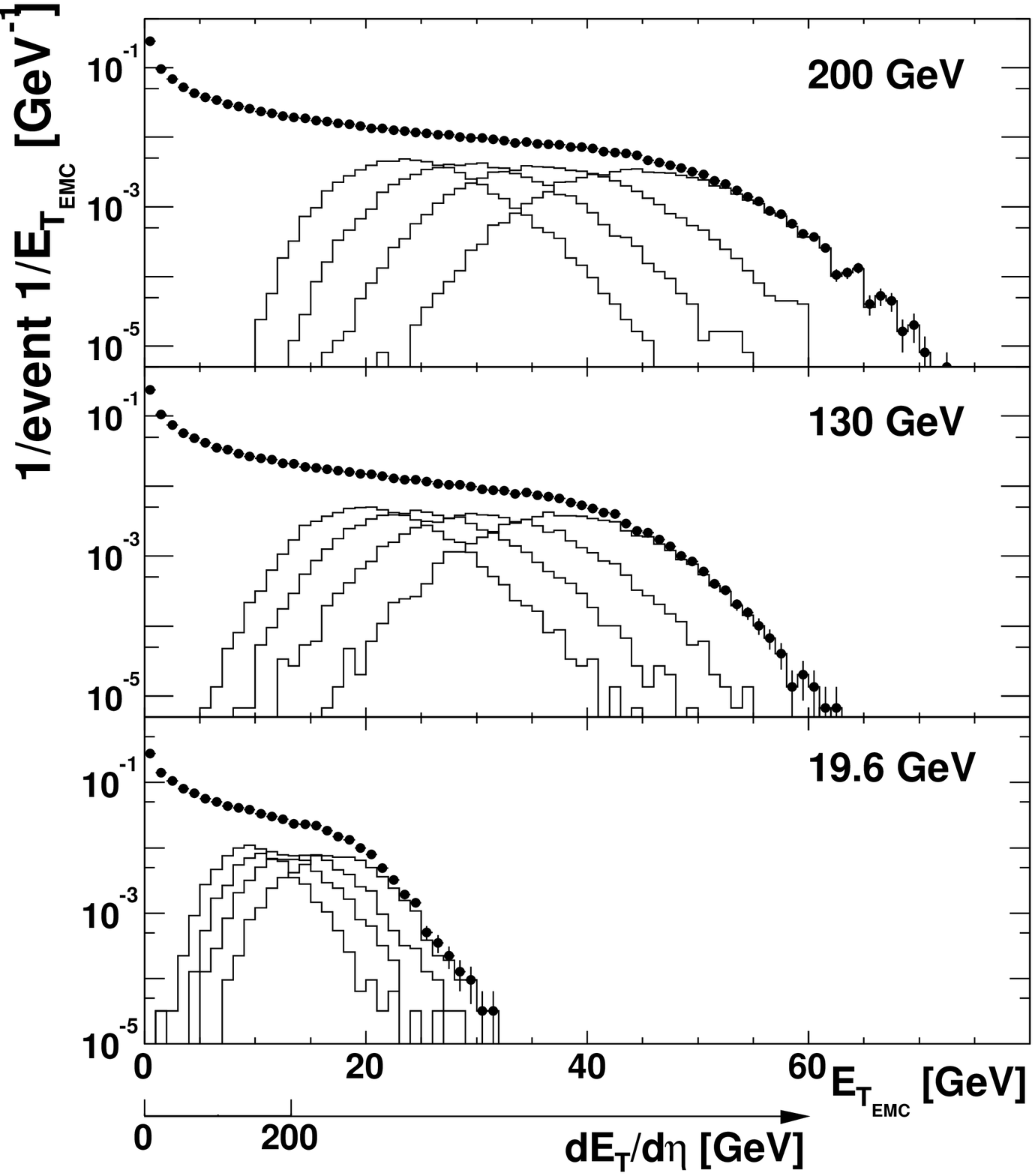}
\includegraphics[width=0.49\linewidth]{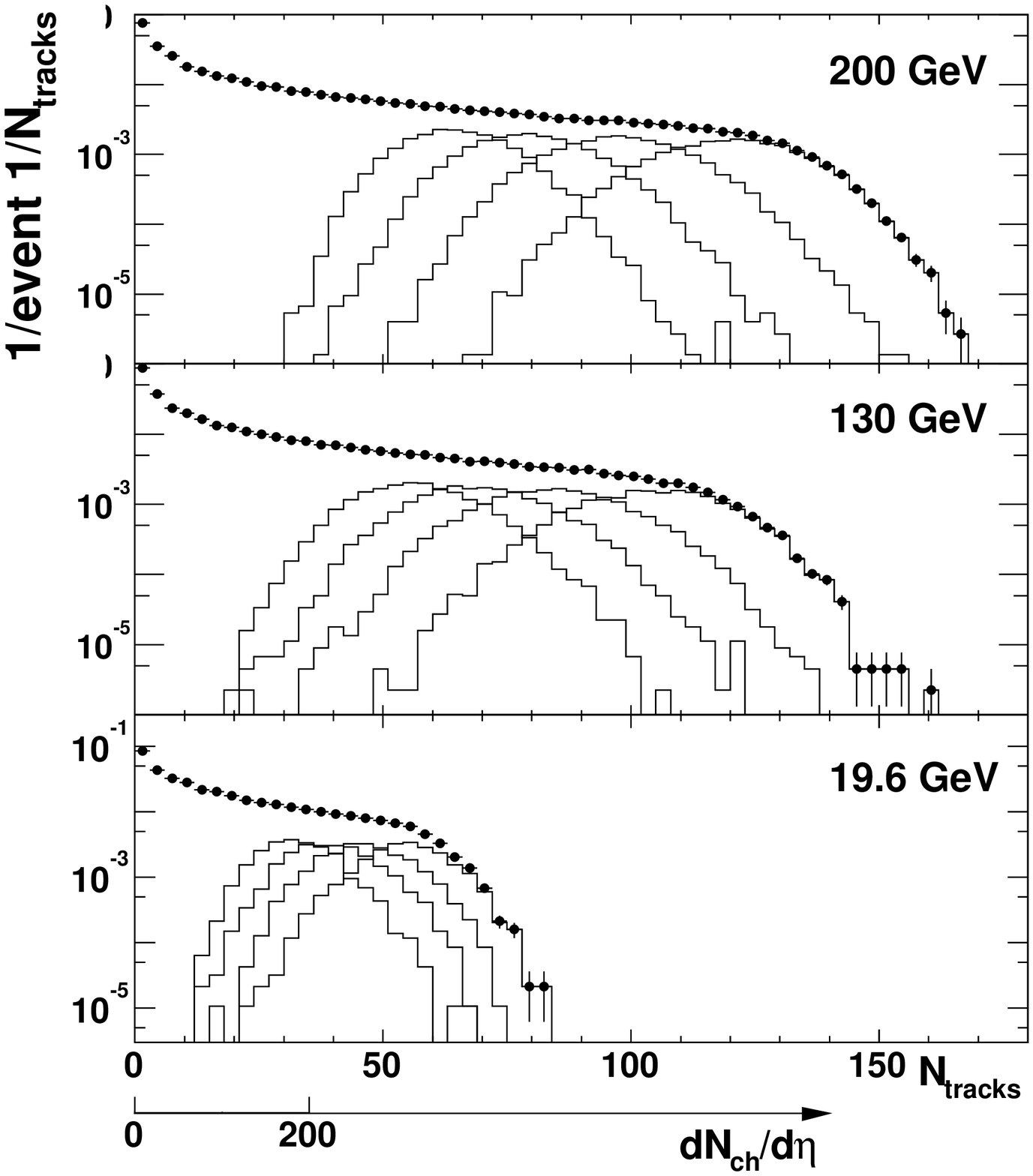}
\caption{The distribution of the raw \Et in two EMCal sectors (left) and the number of 
tracks in the east arm of the PHENIX detector (right) per MB trigger, measured 
at three energies. The lower axis corresponds to mid-rapidity values of 
\dEt and \dNch respectively. Distributions of the four 5\% most central bins 
are also shown in each plot.\label{fig:results_raw}}
\end{figure}

The distribution of the raw transverse energy, $E_{T_{EMC}}$, into the fiducial 
aperture of two EMCal sectors is shown in the left three panels of 
Fig.~\ref{fig:results_raw} for three RHIC energies \sqn = 19.6, 130 and \2. 
The lower scale corresponds to the fully corrected \Et normalized to one unit of 
pseudorapidity and full azimuthal acceptance. The lower axis in the plot is not labeled beyond \2 
to avoid confusion between the true shape of the \dEt distribution 
and \Et as measured using the limited acceptance of two EMCal sectors.

Two EMCal sectors each with azimuthal coverage $\Delta \phi = 27^{o}$ 
were used to make a measurement at \sqn = \3. At other energies we used five EMCal sectors.
Results obtained with different numbers of sectors at the same energy are consistent within 1.5\%.

The right three panels in Fig.~\ref{fig:results_raw} show the number of tracks 
reconstructed in the east arm of the PHENIX detector after the background 
subtraction and all other corrections. The lower axis corresponds to measured 
distributions normalized to one unit of pseudorapidity and full azimuthal 
acceptance. For a similar reason as for the \Et, the lower axis is not labeled above \2 
in $dN_{ch}/d\eta$.

For the \Nch measurements at \sqn = \3, only the east arm was used, while for the other 
energies the measurements were made using both PHENIX central arms. The 
results obtained with two arms at \sqn= 200, 62.4, and \1 are consistent with each 
other within 1.5\%.

The distributions shown in Fig.~\ref{fig:results_raw} have a characteristic shape 
with a sharp peak that corresponds to the most peripheral events. Missing events caused by the finite 
MB trigger efficiency in peripheral events 
would make this peak even sharper than measured. 
The plateau in all distributions corresponds to mid-central 
events and the fall-off to the most central $Au+Au$ events. The shape of the 
curves in Fig.~\ref{fig:results_raw} in the fall-off region is a product of the 
intrinsic fluctuations of the measured quantities and the limited acceptance of the detector.

The distributions for the four most central bins 0\%-5\% to 15\%-20\% are also shown 
in each panel. The centroids of these distributions are used to calculate the 
centrality dependence of \Et and $N_{ch}$~\footnote{All plotted and 
quoted numbers correspond to the centrality bin-by-bin average values or ratios 
of corresponding averages.}. The statistical uncertainty of the mean values (less than or 
about 1\%) determined by the width of the distributions are small because of 
the large size of the event samples.

\begin{figure}[h]
\includegraphics[width=0.49\linewidth]{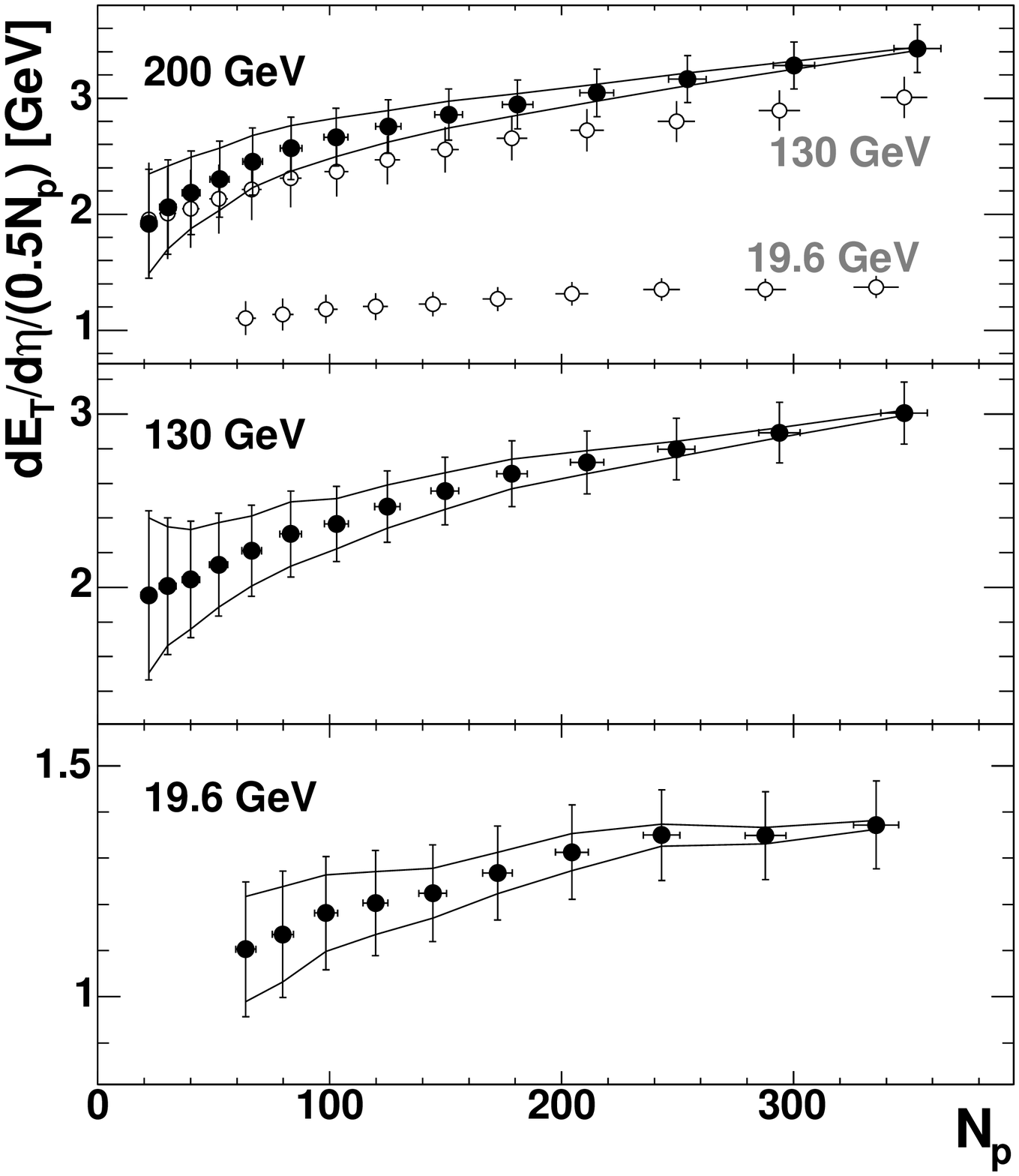}
\includegraphics[width=0.49\linewidth]{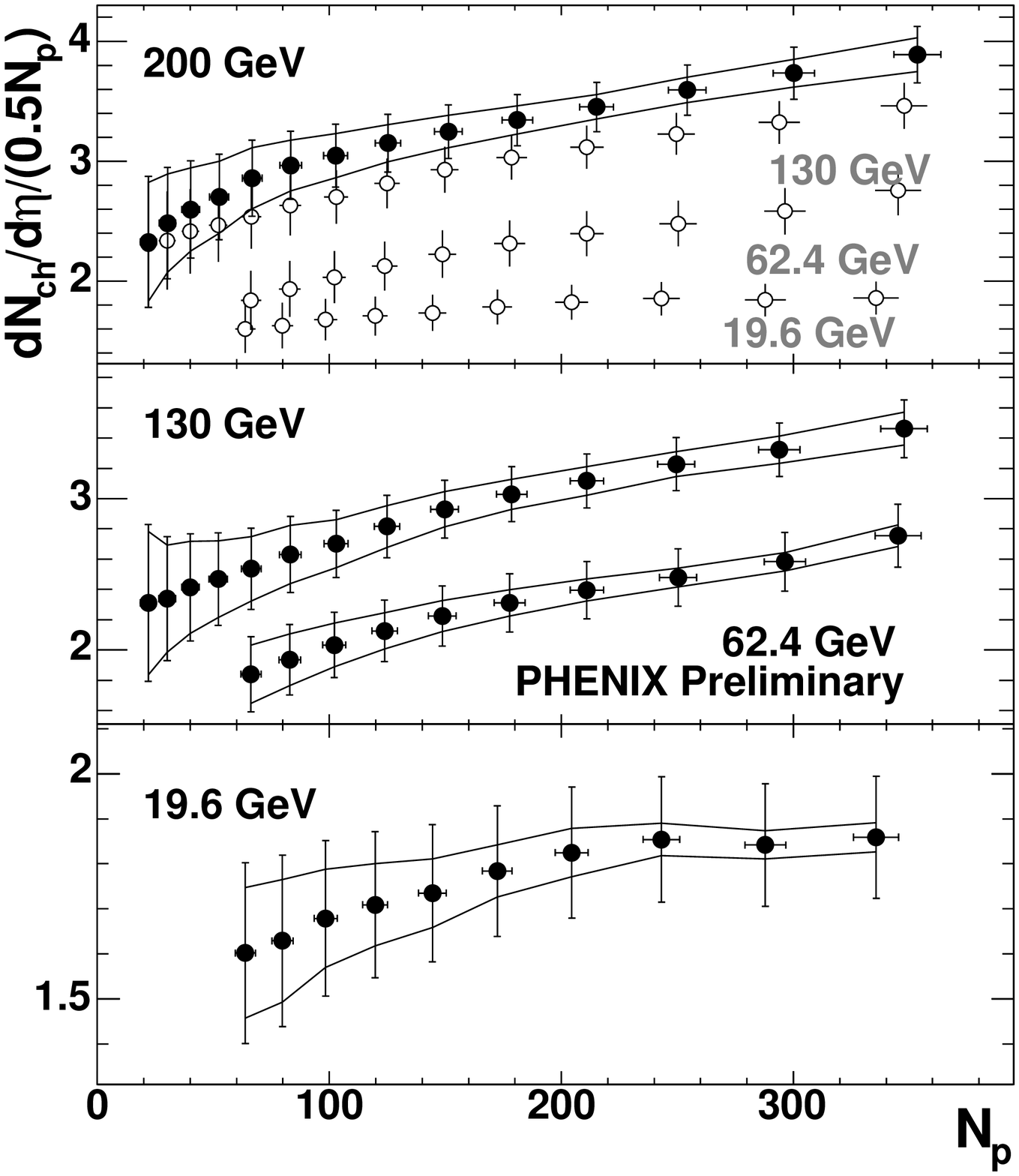}
\caption{\dEt (left) and \dNch (right) divided by the number of participant pairs at 
different RHIC energies. Errors shown with vertical bars are full systematic 
errors. Lines show the part of the systematic error that allows bending or 
inclination of the points. Horizontal errors denote the uncertainty in 
determination of \Nps. PHENIX preliminary result for \Nch at \sqns~=~\6 is shown 
in the right panel.\label{fig:results_vs_npart}}
\end{figure}

The magnitude of \dEt at mid-rapidity 
divided by the number of participant pairs as a function of \Np is shown in Fig.~\ref{fig:results_vs_npart}.
The right three panels show the same ratio for \dNch at three RHIC energies.

The horizontal errors correspond to the uncertainty in \Nps, determined within 
the framework of the Monte Carlo Glauber model. 
The vertical bars show the full systematic errors of the 
measurements added quadratically to the errors of \Nps. 
The lines denote the corridor in which the points can be inclined or bent. 
The statistical errors are smaller than the size of the markers. The upper panel 
also shows the results of the two lower panels with open markers for 
comparison. 

An important result for Fig.~\ref{fig:results_vs_npart} is the consistency evident 
in the behavior of the centrality curves of \Et shown on the left 
and \Nch shown on the right for all measured energies. Both values 
demonstrate an increase from peripheral (65\%-70\%) to the most central events by (50-70)\%
at RHIC energies \sqns=\3 and \2. For the lowest RHIC energy (\sqns=\1) this 
increase is at the level of systematic uncertainties of the measurement. 
One can note that results from PHOBOS~\cite{phobos_total},
show that the total charged particle multiplicity is proportional to \Np
while the multiplicity at mid-rapidity over \Np increases with \Nps,
indicating that the pseudorapidity distribution gets more narrow for
central events. Figure~\ref{fig:results_vs_npart} also show the PHENIX preliminary result for 
\Nch at \sqns~=~\6 added to the central panel.

\begin{figure}[h]
\includegraphics[width=0.49\linewidth]{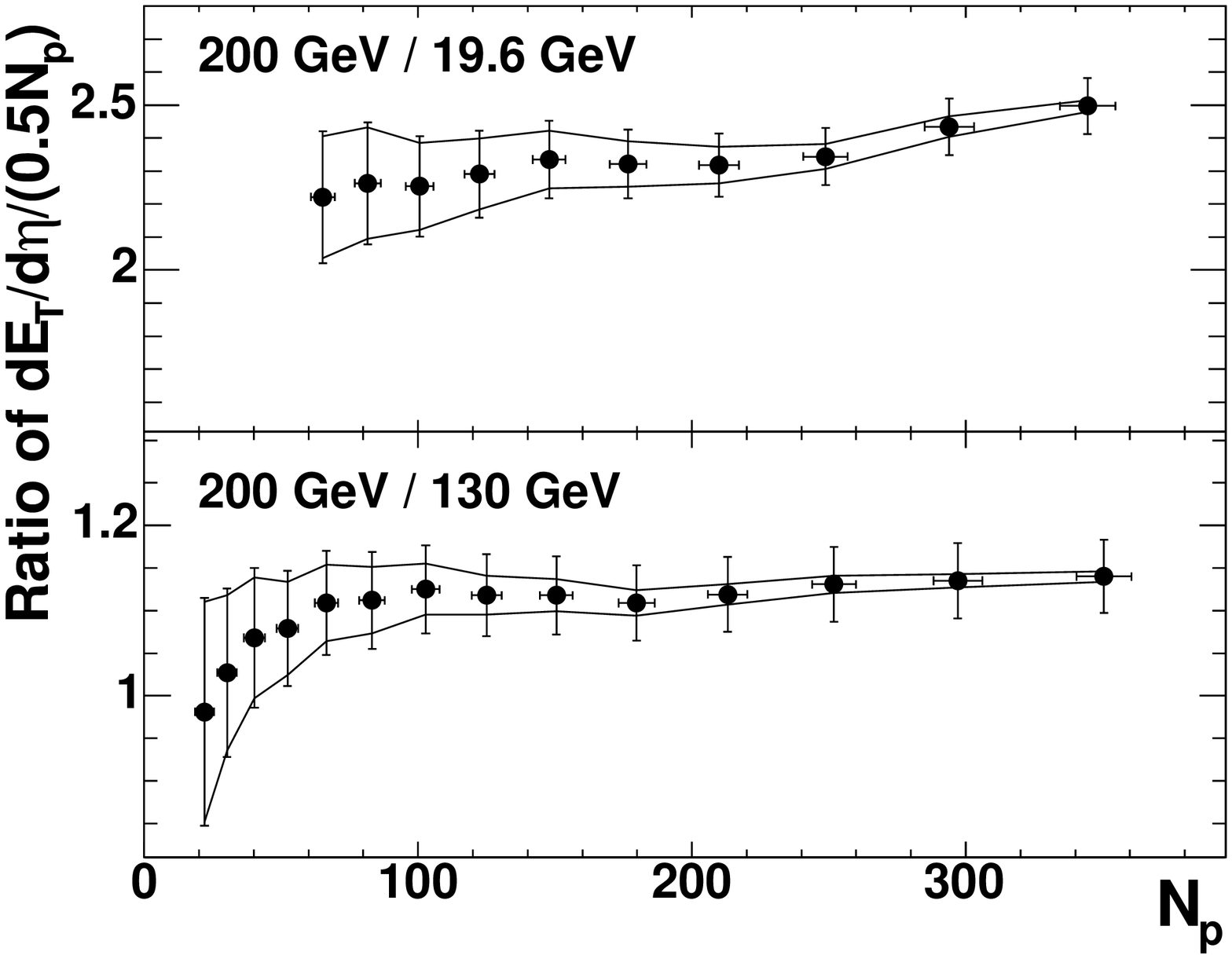}
\includegraphics[width=0.49\linewidth]{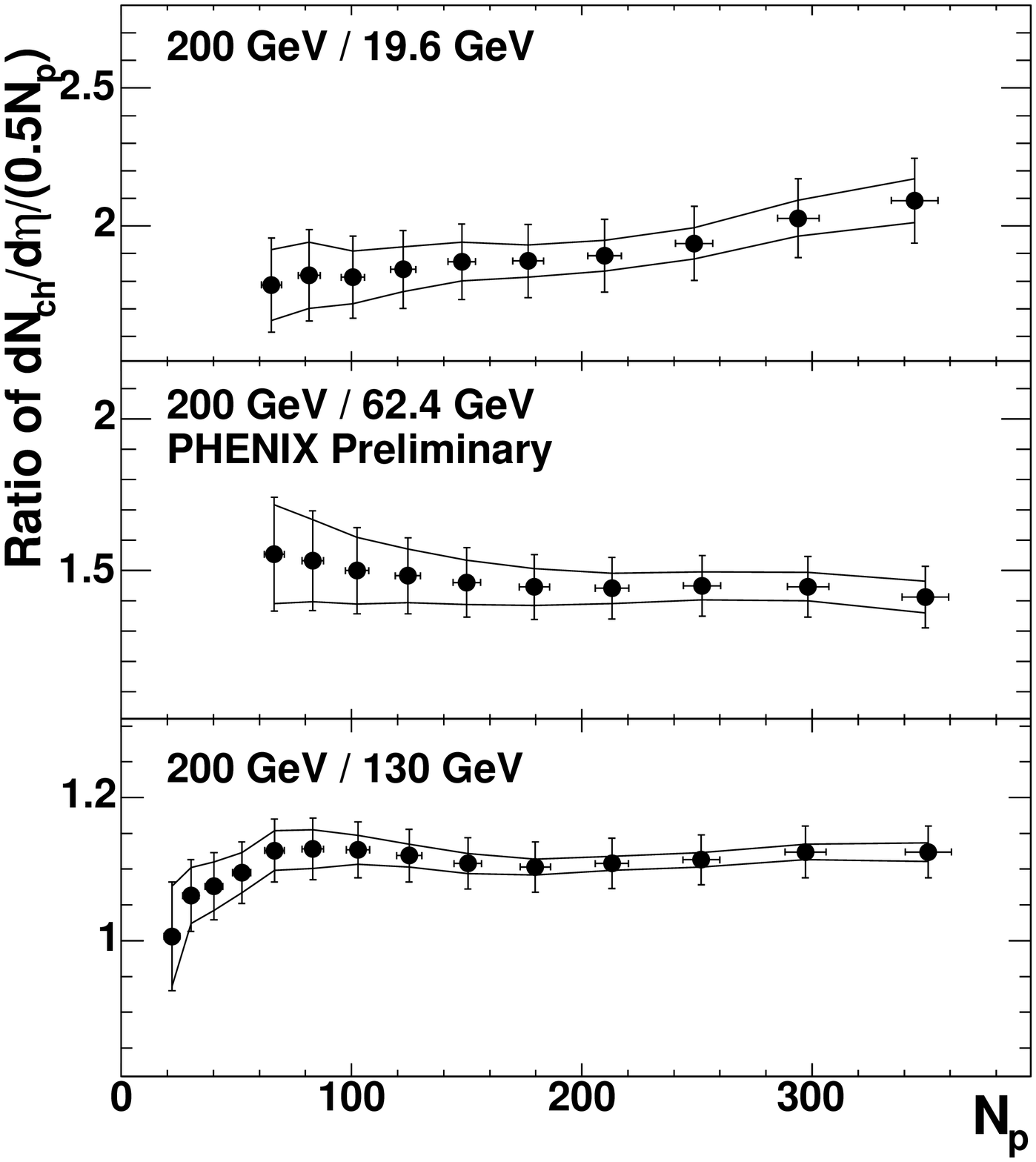}
\caption{Ratios of \dEt (left) and \dNch (right) measured at different RHIC 
energies. The errors shown with vertical bars 
are the full systematic errors. Lines show that part of the systematic error 
that
allows bending or inclination of the points. The horizontal errors denote 
the uncertainty in the determination of \Nps.\label{fig:ratios_vs_npart}}
\end{figure}

The ratios of the \dEt and \dNch per participant pair measured at different 
RHIC energies are shown in Fig.~\ref{fig:ratios_vs_npart}. 
In these ratios some common systematic errors cancel.

The increase in the \Et production between \1 and \2 (with an average 
factor of 2.3) is larger than for \Nch (with average factor of 1.9). This is consistent
 with an increase in the particle production per participant common to both 
\Et and \Nch and a $\sim$20\% increase in \mt of produced particles contributing 
to the \Et parameter only.

The ratio of \2/\1 shows an increase from peripheral to central events; 
however the increase is marginal in comparison to the systematic errors of the 
measurement.

The ratio of \2/\3 is flat above \Np $\sim$ 80 and is equal to $1.140\pm0.043$ for 
\Et and $1.126\pm0.036$ for \Nch in the most central bin. A rather sharp 
increase in the ratios of both quantities between \Np=22 and 83 in the ratios of both quantities 
is still at the level of systematic uncertainties.

The preliminary ratio of \2/\6 although shows the decrease with \Np which is within the systematic errors. 
The increase in the particle production between these two energies is about 40\%.

\begin{figure}[h]
\includegraphics[width=0.49\linewidth]{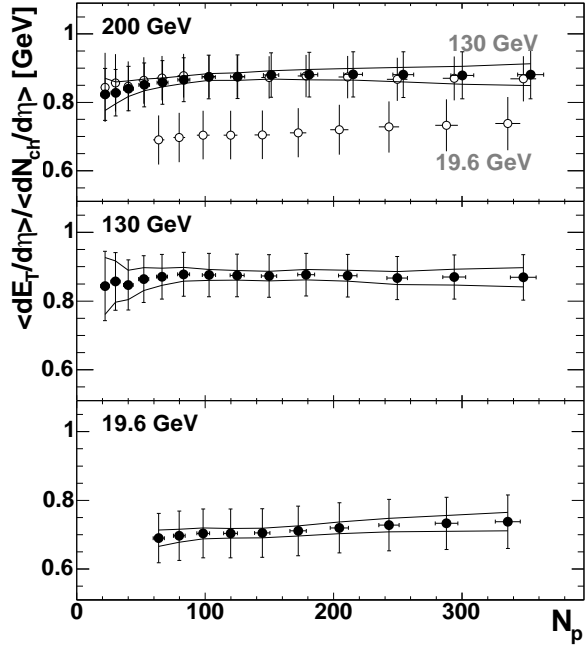}\hspace{2pc}%
\begin{minipage}[b]{18pc}
\caption{\EN vs. \Np at different RHIC energies. The errors shown 
with vertical bars are the full systematic errors. Lines show that part of 
the systematic error that allows bending or inclination of the points. The 
horizontal errors denote the uncertainty in the determination 
of \Nps.\label{fig:etra_mult_vs_npart}}
\end{minipage}
\end{figure}

The ratio of the transverse energy and charged particle multiplicity at mid-rapidity as 
a function of centrality is shown in Fig.~\ref{fig:etra_mult_vs_npart} for 
the three energies. The upper plot also shows the results 
displayed in the lower panels for comparison.

The ratio $E_{T}/N_{ch}$~\footnote{\EN is used as a shortcut for 
$\langle dE_{T}/d\eta \rangle / \langle dN_{ch}/d\eta \rangle$ at $\eta=0$ in C.M.S..}, sometimes 
called the ``Global Barometric Observable'', triggered considerable discussion~\cite{gulash,raju}.
It is related to the \mt of the produced 
particles and is observed to be almost independent of centrality and 
incident energy of the collisions within the systematic errors of the previous 
measurements. The present paper forges a direct link between 
the highest SPS and lowest RHIC energies, making a more quantitative study of 
\EN possible.

The results presented in Fig.~\ref{fig:etra_mult_vs_npart} 
show that the centrality dependence of \EN is weak and lies within the systematic errors 
plotted with lines. There is a clear increase in \EN between 
\sqns=\1 and \2. The \sqn dependence of the results is discussed below.

\subsection{Bjorken Energy Density}
The Bjorken energy density \cite{bjorken} can be obtained using 
\begin{equation} 
\epsilon_{Bj} = \frac{1}{A_{\perp} \tau} \frac{dE_T}{dy}, 
\label{eq:Bj}
\end{equation} 
where $\tau$ is the formation time and $A_{\perp}$ is the nuclei 
transverse overlap area. 

The transverse overlap area of two colliding nuclei was estimated 
using a Monte Carlo Glauber model $A_{\perp} \sim \sigma_x \sigma_y$, 
where $\sigma_x$ and $\sigma_y$ are the widths of $x$ and $y$ position 
distributions of the participating nucleons in the transverse plane. 
The normalization to $\pi R^2$, 
where $R$ is the sum of $r_n$ and $d$ parameters in a Woods-Saxon 
parameterization (see e.g.:~\cite{ppg19}), was done for the most central 
collisions at the impact parameter $b=0$.

For the transformation from $dE_T/d\eta|_{\eta=0}$ to $dE_T/dy|_{y=0}$, 
a scale factor of $1.25 \pm 0.05$ was used, see~\cite{ppg19} for the details of conversion procedure.

\begin{figure}[h]
\includegraphics[width=0.49\linewidth]{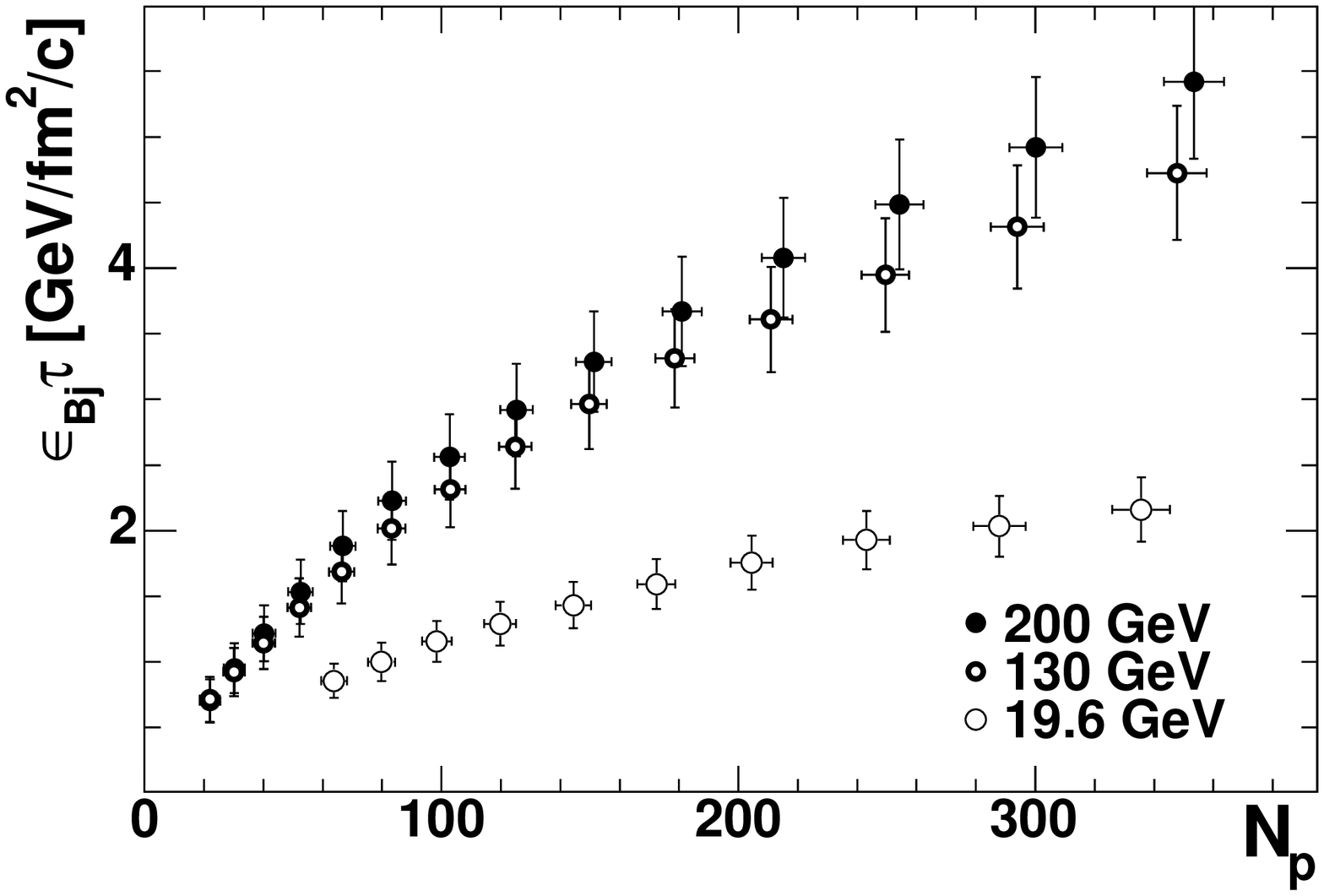}
\includegraphics[width=0.49\linewidth]{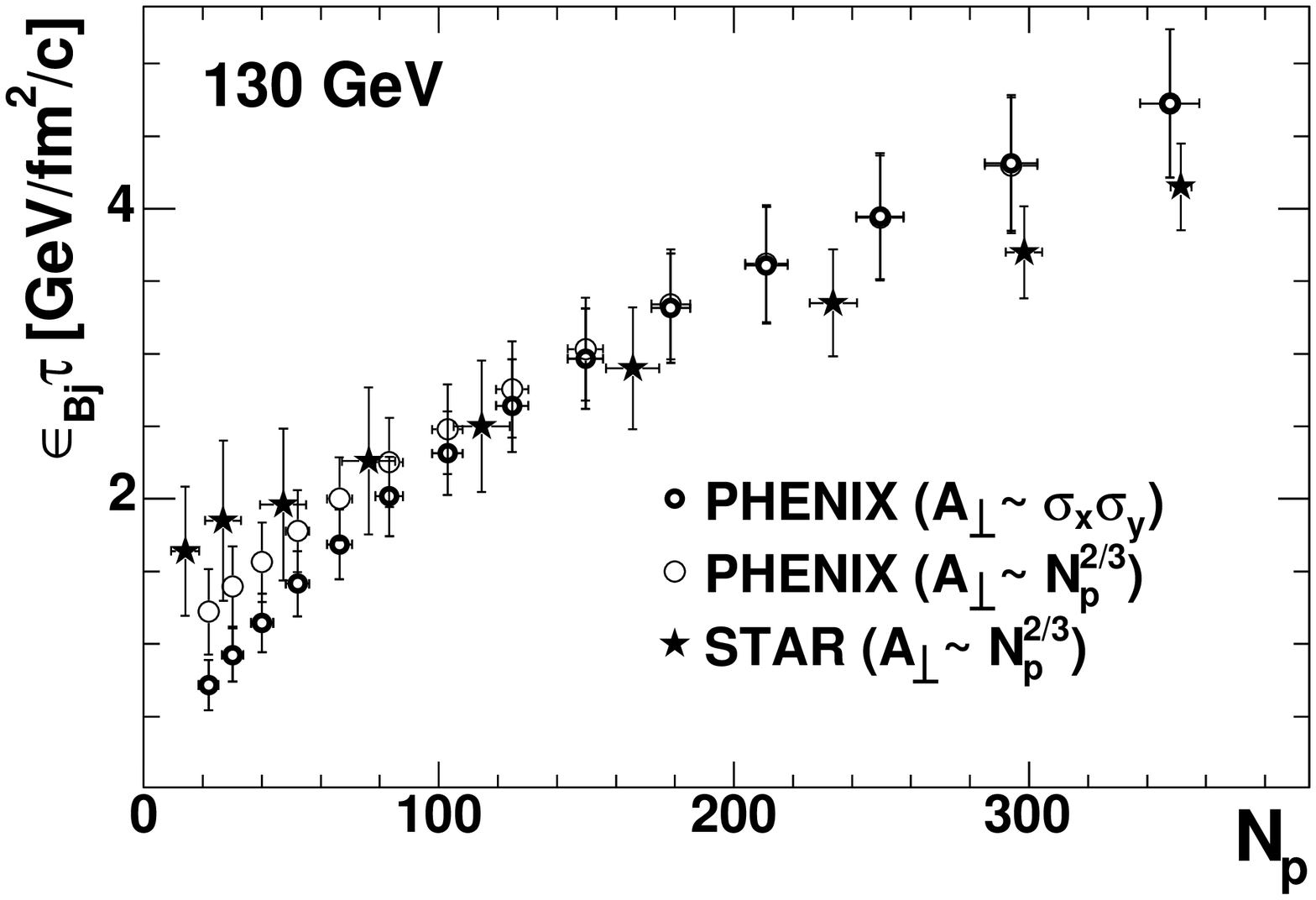}
\caption{$\epsilon_{Bj} \cdot \tau$ deduced from the PHENIX data at three RHIC energies (left) 
and using different estimates of the nuclear transverse overlap at \sqn=\3 (right). 
\label{fig:ebj}}
\end{figure}

The Bjorken energy density for three RHIC energies is plotted in the left panel of Fig.~\ref{fig:ebj}.
For the 5\% most central collisions, $\epsilon_{Bj} \cdot \tau$ was 
$2.2 \pm 0.2$, $4.7 \pm 0.5$ and $5.4 \pm 0.6$ GeV/($fm^2 \cdot c$) for 
\sqn=19.6, 130 and 200 GeV, respectively. These values increase by 
2\%, 4\% and 5\%, respectively, for the maximal $N_{part}$=394, as obtained 
from extrapolation of PHENIX data points. There is a factor 
of 2.6 increase between the ``SPS''-like energy (\sqn=\1) and the top RHIC energy \sqn=\2.
The comparison of the only published 
$\epsilon_{Bj}$=3.2 GeV/$fm^3$ at SPS~\cite{na49_4} 
and top RHIC energies, assuming the same $\tau$=1~fm/c, reveals an increase in 
energy density by a factor of only 1.8, which may come from an overestimation 
in the SPS measurement, as shown in the left panel of Fig.~\ref{fig:sqn} below.

Another approach is used by STAR in~\cite{star_bj} for the 
estimate of the transverse overlap area of the two nuclei $A_{\perp} \sim N_{p}^{2/3}$ 
in Eq.~\ref{eq:Bj}. This approach accounts only for the common area of colliding 
nucleons, not nuclei. The results differ only in the peripheral bins 
as shown in the right panel of Fig.~\ref{fig:ebj}. For a comparison, the same 
panel shows the result obtained by STAR which agrees with PHENIX result 
within systematic errors, displaying a smaller increase of the energy 
density with $N_p$.

Can we derive an estimate for $\tau$? We might say, on general quantum mechanical grounds,
that a particle of full energy close to $m_{T}$ can be
considered to have ``formed'' after a time $t=\hbar/m_{T}$ since its creation in that frame.

To estimate the average transverse mass, we use the final-state
\EN results presented in Fig.~\ref{fig:etra_mult_vs_npart}. As it was mentioned above, this value is 
basically independent of centrality and changes from 0.74 GeV to 0.88 GeV for \sqn from 19.6 to 200~GeV.
Such estimate would provide us with the formation time $\tau$ which is almost the same for all initial 
conditions of the collisions at RHIC.

An approximate factor of 2/3 is used to account for the total number of particles making the estimate of
$\langle m_{T} \rangle \simeq$0.57~GeV at \sqn=\2 and \3 and $\langle m_{T} \rangle \simeq$0.50~GeV at \sqn=19.5~GeV.
This results in a formation time of $\tau \simeq$0.35~$fm/c$ and 0.40~$fm/c$ at 
these energies respectively. Then for the 5\% most central events the estimate of the Bjorken 
energy at the three RHIC energies are 5.5~GeV/$fm^3$, 13~GeV/$fm^3$ and 15~GeV/$fm^3$.

\subsection{Comparison to other measurements}
Several factors complicate the comparison between the results of PHENIX and 
the results of other experiments. AGS and SPS data were taken in the Laboratory (Lab.) system 
while the RHIC data are in the Center of Mass (C.M.S.) system. Since
$\eta$ and \Et are not boost invariant quantities, the data should be converted into 
the same coordinate system. Some experiments provide a complete 
set of identified particle spectra from which information about \Et and 
\Nch can be deduced, while other experiments require additional assumptions to extract \dEt and \dNch
from their results. Publication~\cite{ppg19} discusses it in details

\begin{figure}[h]
\includegraphics[width=0.49\linewidth]{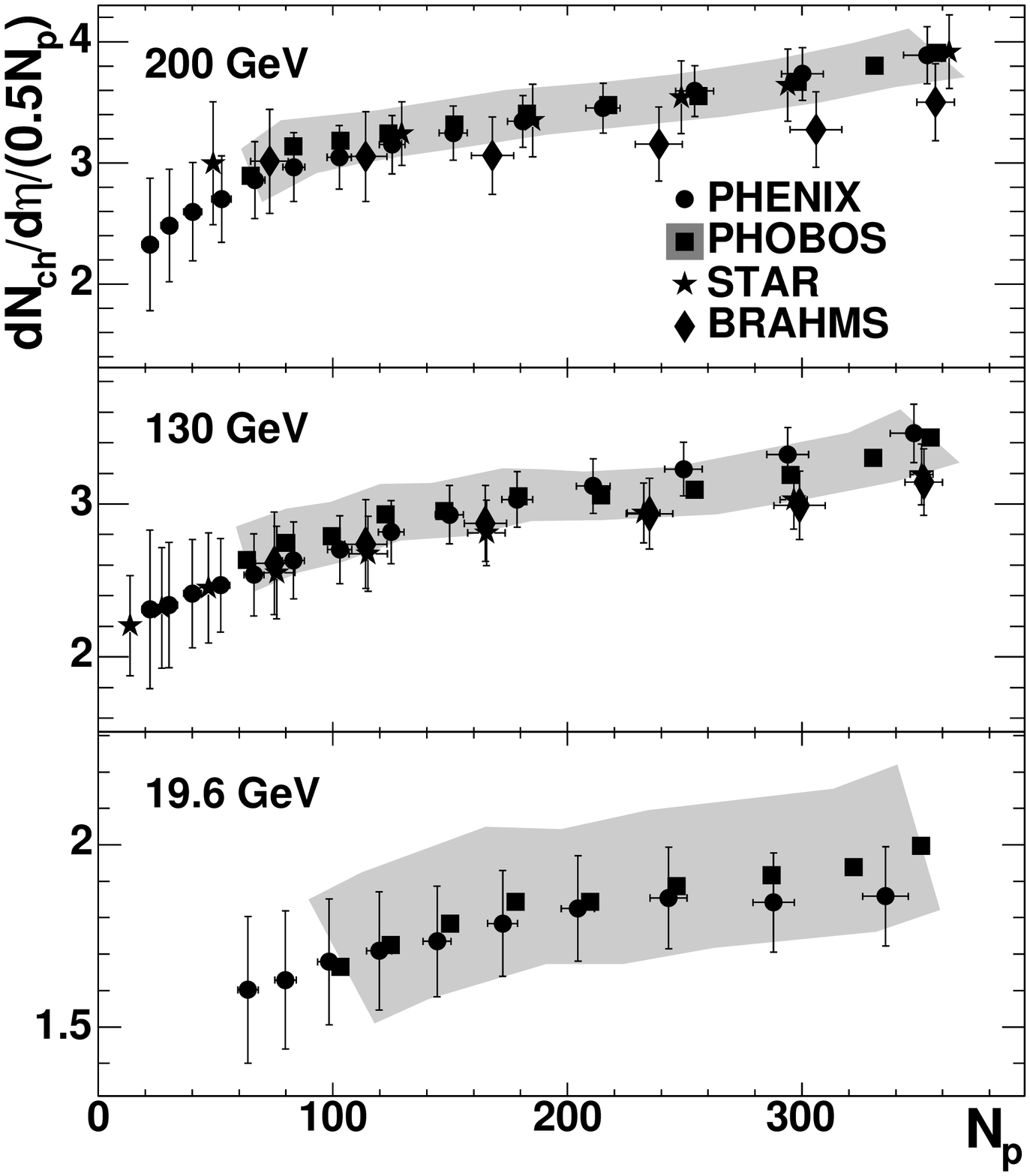}
\includegraphics[width=0.49\linewidth]{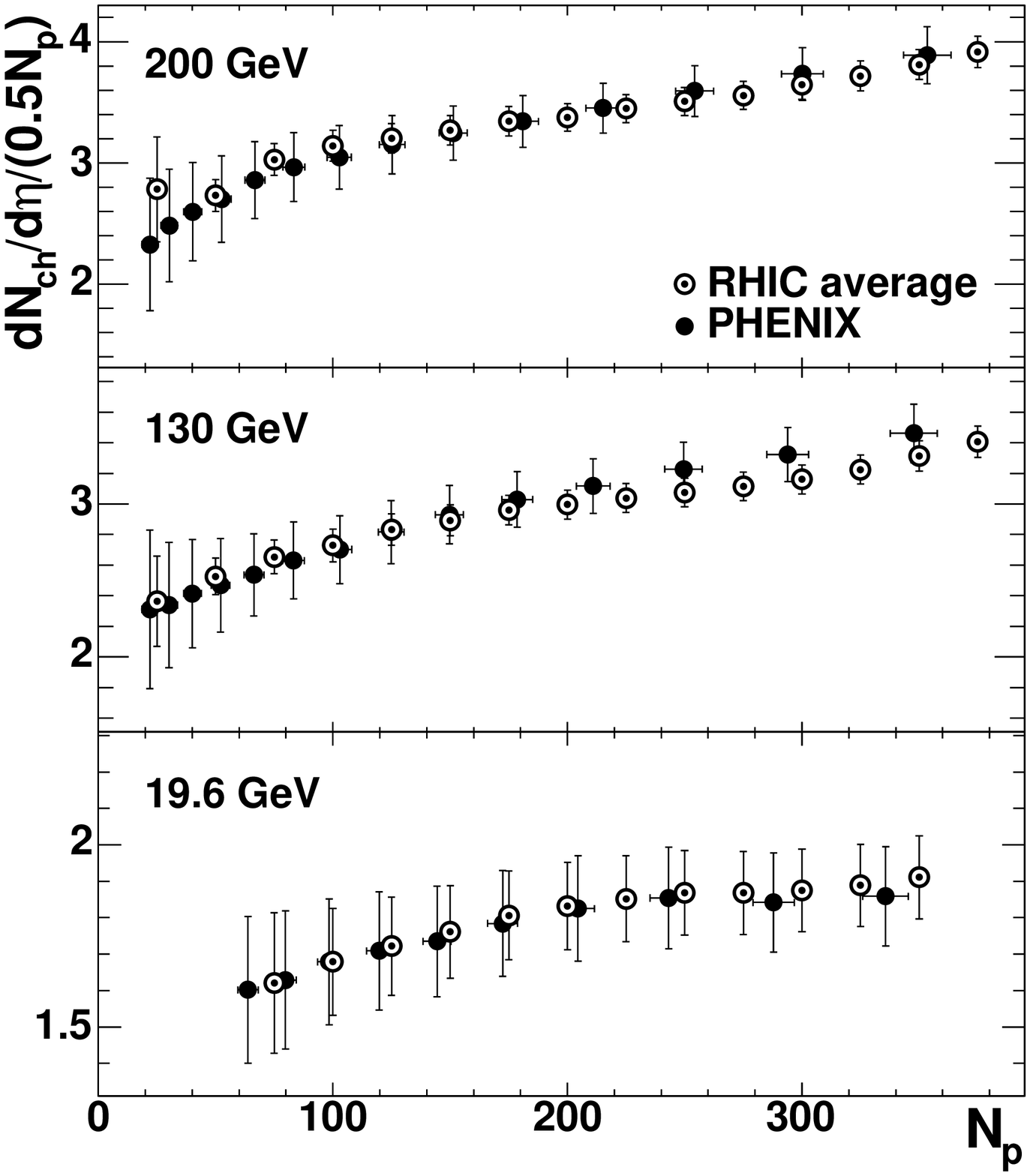}
\caption{Left panel: \dNch per pair of \Np measured by the four RHIC experiments at 
different energies. The shaded area is the PHOBOS systematic error. Right panel: 
RHIC average values (including PHENIX) compared to the PHENIX results.\label{fig:comp_rhic}}
\end{figure}

The PHENIX results for \Nch are compared to the data available from the other 
RHIC experiments. This comparison is shown in the left panels of 
Fig.~\ref{fig:comp_rhic}.

There is good agreement between the results of BRAHMS~\cite{brahms1,brahms2}, 
PHENIX, PHOBOS~\cite{phobos1,phobos2,phobos3} and STAR~\cite{star1,star2} using 
\Np based on a Monte-Carlo Glauber model. This agreement is very impressive because 
all four experiments use different apparatuses and techniques to measure the 
charged particle production. The systematic errors of all results are uncorrelated, 
except for errors due to the same Glauber model which are small.
That makes it possible to calculate the RHIC average and reduce 
the systematic uncertainty. The averaged results from all four RHIC experiments 
are plotted in the right panel of Fig.~\ref{fig:comp_rhic}. The procedure of the 
calculating the average is the same as used in~\cite{PDG} and is also explained 
in~\cite{ppg19}.

\begin{figure}[h]
\includegraphics[width=0.49\linewidth]{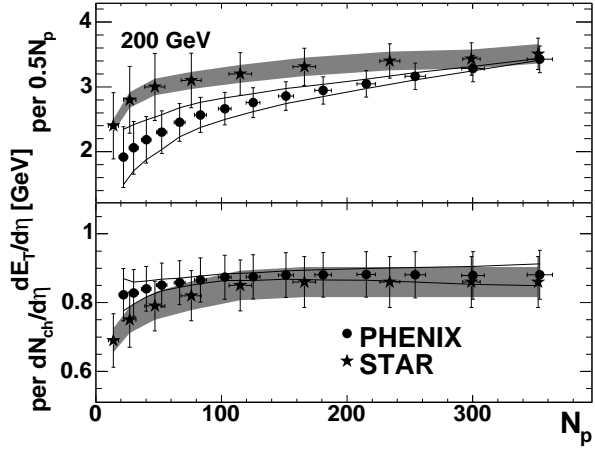}\hspace{2pc}%
\begin{minipage}[b]{18pc}
\caption{\dEt divided by the number of \Np pairs (top) and  \EN (bottom) measured 
by the PHENIX and STAR~\cite{star_etra} experiments at \sqn=200 GeV. PHENIX systematic 
errors are explained in the text. The shaded area is the STAR systematic scaling error.\label{fig:etra_star}}
\end{minipage}
\end{figure}

Figure~\ref{fig:etra_star} compares $E_T$ results from the PHENIX and STAR~\cite{star_etra}
experiments. The results are consistent for all centralities within systematic errors, though 
STAR \dEt per participant pair as shown in the upper panel has a smaller slope for $N_p$ 
in going from from semi-peripheral to central collisions and \EN shown in the lower panel is consistent 
for all \Nps.

\begin{figure}[h]
\includegraphics[width=0.49\linewidth]{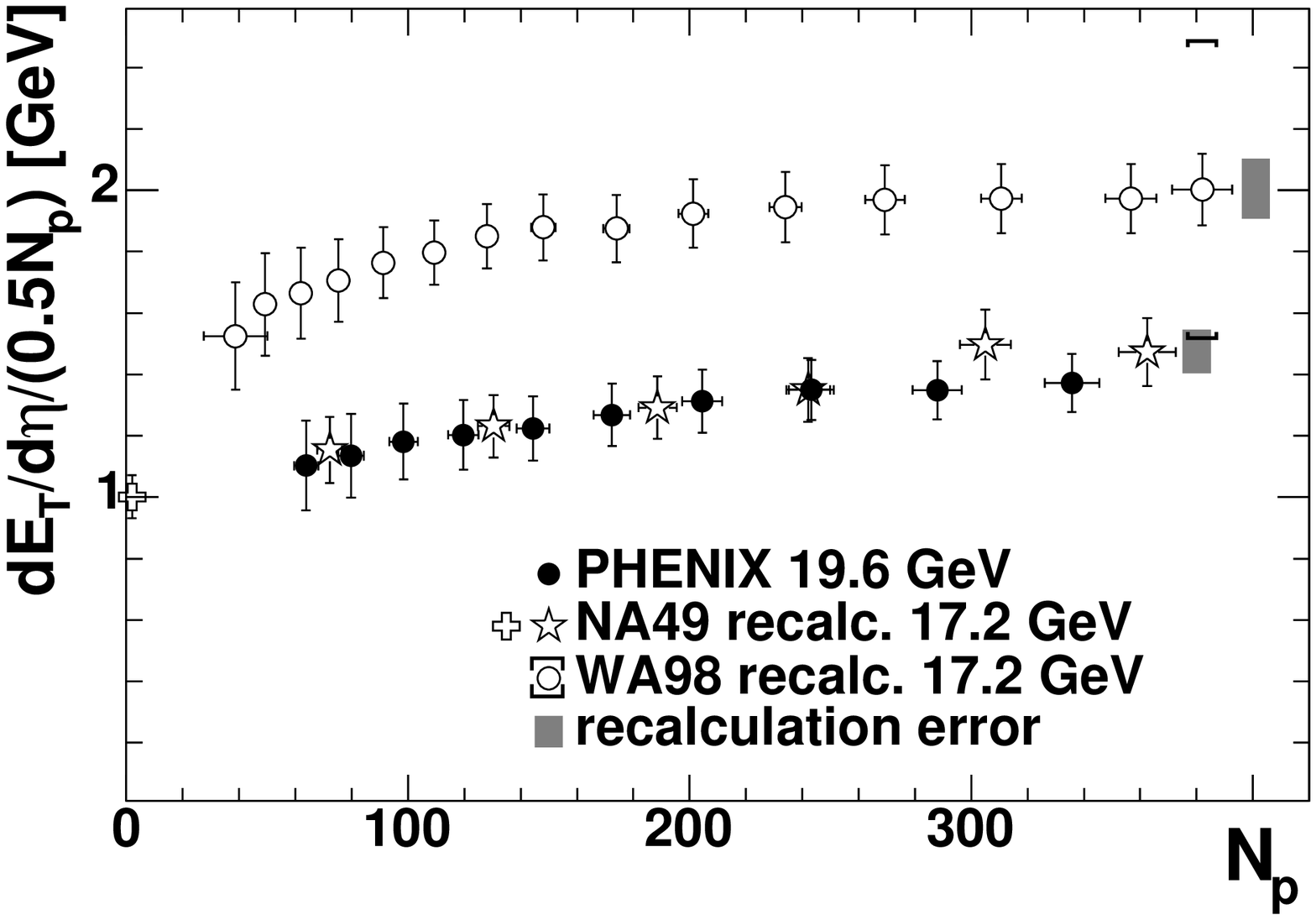}
\includegraphics[width=0.49\linewidth]{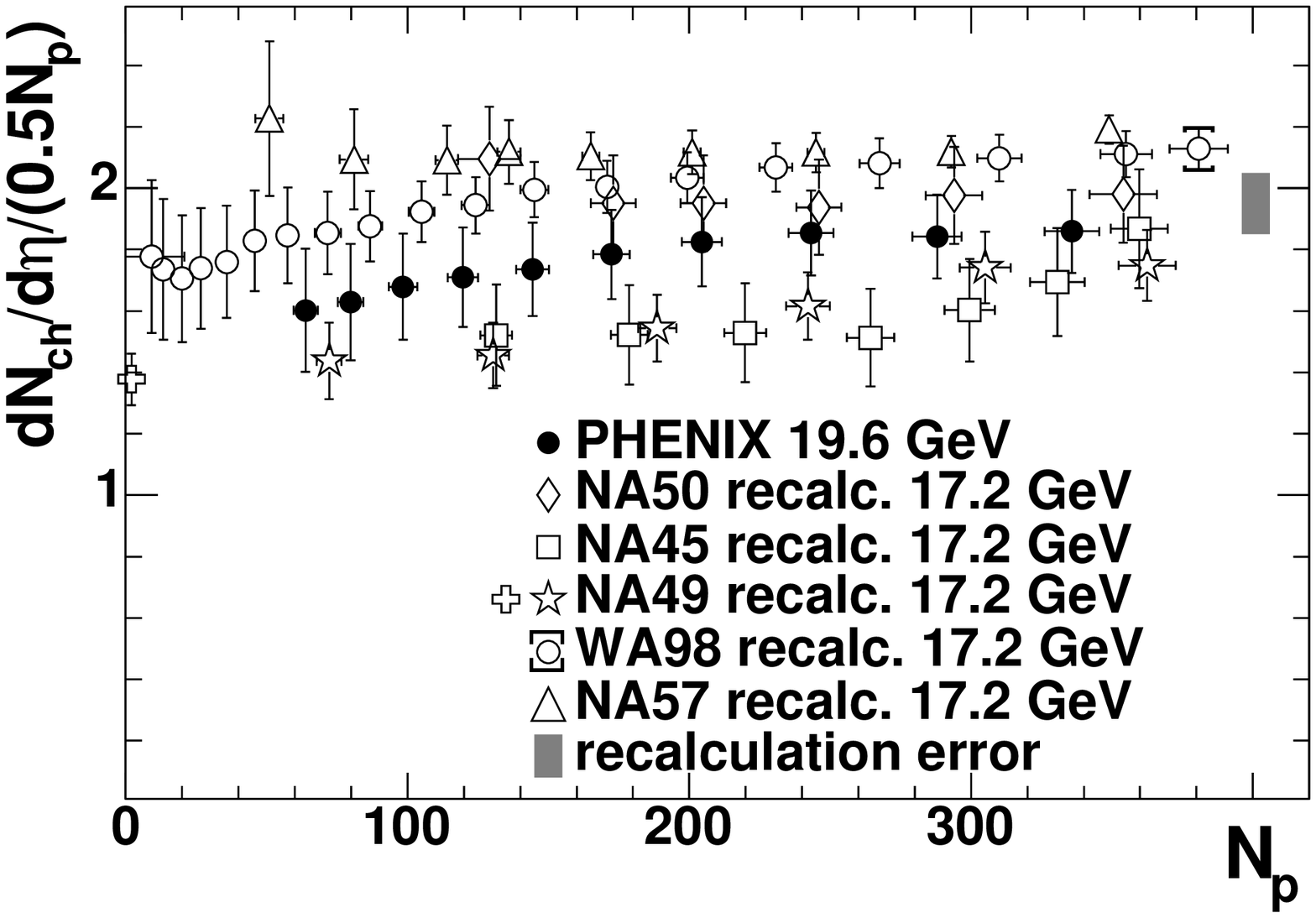}
\caption{\dEt (left) and \dNch (right) divided by the number of \Np pairs 
measured by PHENIX at \sqns=\1 (solid markers) and recalculated from 
the results of the SPS experiments at the highest energy \sqns=\7 (open markers). 
The $p+p$ result of NA49 is marked with an open cross.
\label{fig:comp_sps}}
\end{figure}

The RHIC run at \sqns=\1 allows us to make a connection between 
RHIC and SPS data. The highest SPS energy of 158A~GeV corresponds to 
\sqns~=~\7 in the C.M.S., making a direct comparison of RHIC and SPS results 
possible. This comparison is shown in Fig.~\ref{fig:comp_sps}. 
The SPS data is taken from~\cite{na49_1,na49_2,na49_3,wa98,ceres_1,na50_1}
details of the data compilation are explained in \cite{ppg19}.

Several comments should be made about this comparison. For both measured 
parameters the PHENIX results and the SPS results agree. The WA98 results 
are systematically higher, especially for $dE_{T}/d\eta$. However the WA98 data has 
an additional systematic error common to all points shown for the last bin. 
For \Nch the relative spread of the SPS results is 
larger than for the RHIC results shown in Fig.~\ref{fig:comp_rhic}, 
though overall the \sqns=\7 SPS measurements are consistent with the PHENIX 
result at \sqns~=~19.6 GeV. The NA57 results at this energy and at lower energy 
are published without systematic errors~\cite{na57}, so they cannot be compared 
to other results on the same basis.

\begin{figure}[h]
\includegraphics[width=0.49\linewidth]{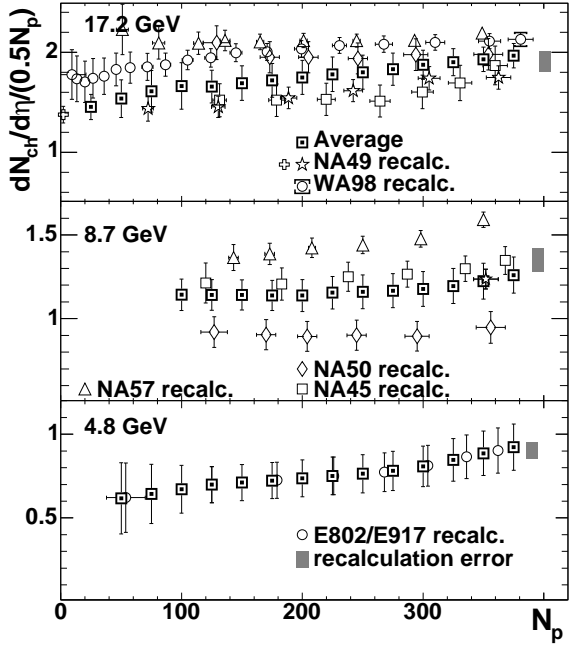}\hspace{2pc}%
\begin{minipage}[b]{18pc}
\caption{\dNch divided by the number of \Np pairs measured by AGS and SPS 
experiments and the average taken at different energies recalculated in the C.M.S.
\label{fig:sps_ags}}
\end{minipage}
\end{figure}

Different SPS and AGS experiments made measurements at lower energies. The 
combined data of AGS, SPS and RHIC provide a complete picture of the 
centrality behavior of \Et and \Nch as a function of the nucleon-nucleon 
energy. The centrality dependence of \dNch at mid-rapidity measured at 
\sqns=4.8, 8.7 and \7 by different experiments is shown in Fig.~\ref{fig:sps_ags}. 
The data is taken form~\cite{na49_2,na49_3,ceres_3,na50_1,ceres_2,e802_1,e802_2,e802_3,e802_5,e917_4,e802_6} 
and the details of data compilation can be found in~\cite{ppg19}.

At the highest SPS energy the averaging procedure is the same as for RHIC 
energies and weighted experimental errors are scaled with the reduced 
$\chi^{2}$-like factor $S$ 
reaching the value of 1.5 at some points. For the intermediate SPS 
energy \sqns=8.7~GeV, two experiments NA45~\cite{ceres_3} and 
NA50~\cite{na50_1} reported the centrality dependence of \dNch at 
mid-rapidity. The discrepancy in the measurements is approximately three times
the quadratic sum of their systematic errors. However the shapes of the two curves are almost 
the same. NA49 has published results~\cite{na49_2,na49_3} 
which give one point in \dNch at \Np=352. This point favors 
the NA45 result. The NA57 results~\cite{na57} also plotted in the Fig.~\ref{fig:sps_ags} 
demonstrate even larger discrepancy with the NA50 measurements, but these points 
lack the systematic errors, as we mentioned before.

The average centrality curve is produced by taking into 
account the shape of the centrality curves reported by NA45 and NA50 
and the single NA49 point. The errors are scaled with the 
factor $S$, which reaches a value of 2.5 at some points. The AGS results 
are presented with a curve produced from the combined results of the E802/E917 
experiments (see publication~\cite{ppg19}).

The average SPS centrality dependence at \sqns~=~\7 shown in the upper panel 
in Fig.~\ref{fig:sps_ags} and the average curve of the two RHIC 
experiments at \sqns=\1 shown in the lower panel in Fig.~\ref{fig:comp_rhic} 
are very similar. Less than 5\% increase is expected due to the difference in
the incident energies between the highest SPS and the lowest RHIC 
energies (see section~\ref{sec:central_collision} below). 

The average values presented in Figs.~\ref{fig:comp_rhic} 
and \ref{fig:sps_ags} are summarized in Table~\ref{tab:averages}.

\begin{center}
\begin{table}
\caption{ Average values of \dNch/(0.5\Nps) at different \sqns. An 
additional 5\% error should be added to columns \7 through 4.8~GeV for 
the uncertainty related to recalculation to the C.M.S..
\label{tab:averages}}
\centering
\begin{tabular}{cccccccc}
\br
\Np & \2   & \3   & \6     & \1     & \7    & 8.7~GeV   & 4.8~GeV   \\
    & RHIC & RHIC & PHENIX & PHENIX/& SPS   & SPS       & E806/E917 \\
    & aver.& aver.& prelim.& PHOBOS & aver. & aver.     & combined  \\
\mr
375 & 3.92$\pm$0.13 & 3.41$\pm$0.10 & 	            &               & 1.97$\pm$0.12 & 1.26$\pm$0.11 & 0.92$\pm$0.14\\
350 & 3.81$\pm$0.13 & 3.31$\pm$0.10 & 2.77$\pm$0.21 & 1.91$\pm$0.11 & 1.93$\pm$0.12 & 1.22$\pm$0.11 & 0.89$\pm$0.13\\
325 & 3.72$\pm$0.12 & 3.22$\pm$0.10 & 2.68$\pm$0.20 & 1.89$\pm$0.11 & 1.90$\pm$0.14 & 1.20$\pm$0.11 & 0.85$\pm$0.13\\
300 & 3.65$\pm$0.12 & 3.16$\pm$0.10 & 2.60$\pm$0.20 & 1.88$\pm$0.11 & 1.88$\pm$0.15 & 1.18$\pm$0.10 & 0.81$\pm$0.12\\
275 & 3.56$\pm$0.12 & 3.11$\pm$0.09 & 2.54$\pm$0.19 & 1.87$\pm$0.11 & 1.83$\pm$0.16 & 1.17$\pm$0.10 & 0.78$\pm$0.12\\
250 & 3.51$\pm$0.12 & 3.07$\pm$0.09 & 2.48$\pm$0.19 & 1.87$\pm$0.12 & 1.80$\pm$0.17 & 1.16$\pm$0.10 & 0.76$\pm$0.11\\
225 & 3.45$\pm$0.12 & 3.04$\pm$0.10 & 2.43$\pm$0.19 & 1.85$\pm$0.12 & 1.78$\pm$0.17 & 1.16$\pm$0.10 & 0.75$\pm$0.11\\
200 & 3.38$\pm$0.11 & 3.00$\pm$0.09 & 2.37$\pm$0.19 & 1.83$\pm$0.12 & 1.75$\pm$0.17 & 1.14$\pm$0.10 & 0.74$\pm$0.11\\
175 & 3.34$\pm$0.12 & 2.96$\pm$0.10 & 2.30$\pm$0.19 & 1.81$\pm$0.12 & 1.72$\pm$0.17 & 1.14$\pm$0.09 & 0.72$\pm$0.11\\
150 & 3.27$\pm$0.12 & 2.89$\pm$0.10 & 2.23$\pm$0.20 & 1.76$\pm$0.13 & 1.69$\pm$0.17 & 1.14$\pm$0.09 & 0.71$\pm$0.11\\
125 & 3.20$\pm$0.12 & 2.83$\pm$0.10 & 2.13$\pm$0.20 & 1.72$\pm$0.14 & 1.66$\pm$0.16 & 1.14$\pm$0.09 & 0.70$\pm$0.11\\
100 & 3.14$\pm$0.13 & 2.73$\pm$0.11 & 2.02$\pm$0.22 & 1.68$\pm$0.15 & 1.66$\pm$0.23 & 1.14$\pm$0.09 & 0.67$\pm$0.14\\
 75 & 3.03$\pm$0.13 & 2.65$\pm$0.11 & 1.89$\pm$0.24 & 1.62$\pm$0.19 & 1.61$\pm$0.21 &               & 0.64$\pm$0.18\\
 50 & 2.73$\pm$0.13 & 2.53$\pm$0.12 & 	            &               & 1.54$\pm$0.19 &               & 0.63$\pm$0.21\\
 25 & 2.78$\pm$0.43 & 2.36$\pm$0.30 & 	            &               & 1.45$\pm$0.13 &               &              \\
\br				      	     
\end{tabular}			                   
\end{table}
\end{center}
 
\subsection{Dependence on the incident nucleon energy.}
The data compilation made in the previous section allows for
a detailed study of the charged particle production in heavy ion reactions 
at different incident energies of colliding nuclei. Although the data on
transverse energy production is not abundant, a similar comparison can
be made~\cite{phenix_milov,phenix_bazik}.

\subsubsection{Central Collisions}
\label{sec:central_collision}
\begin{figure}[h]
\includegraphics[width=0.49\linewidth]{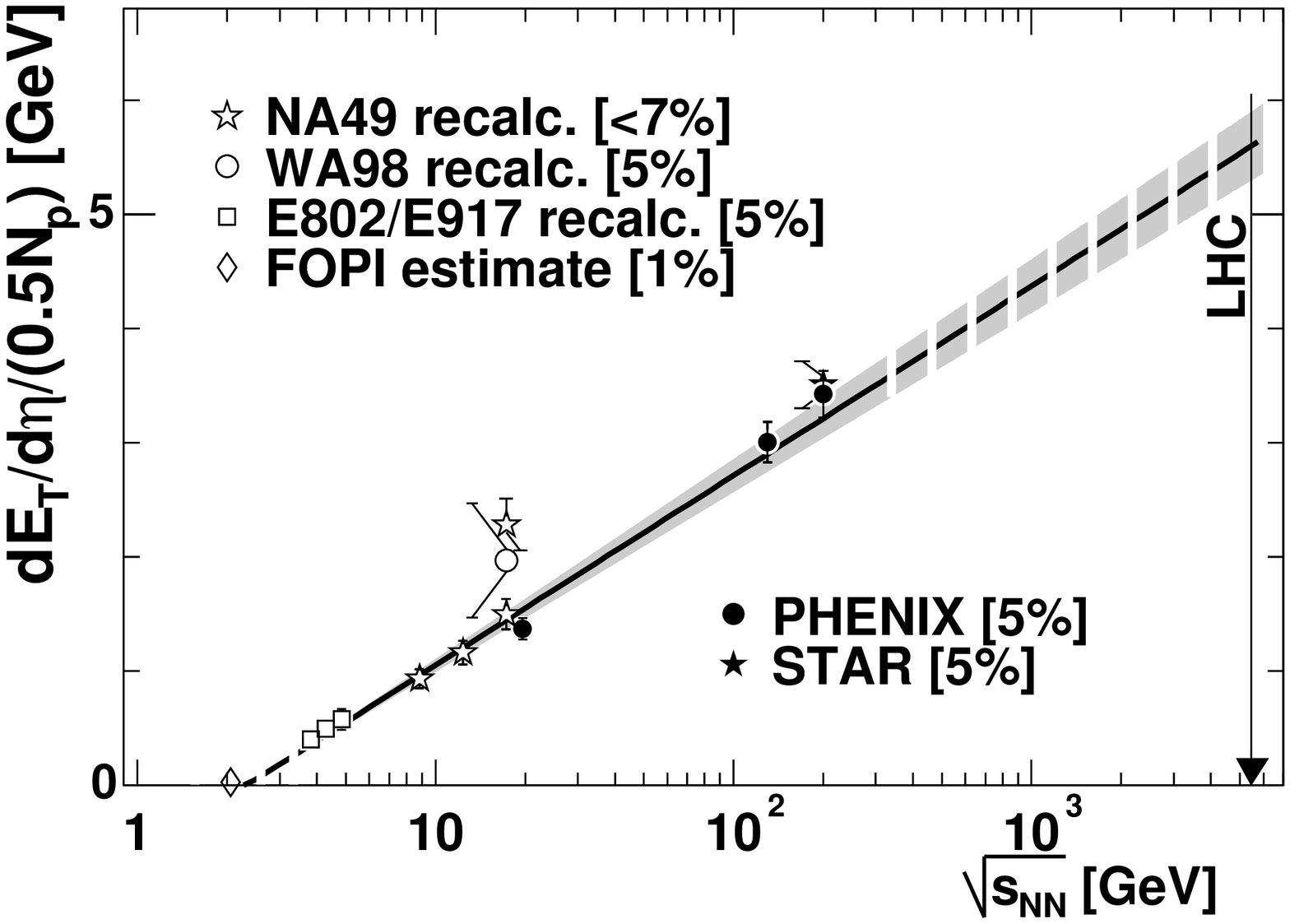}
\includegraphics[width=0.49\linewidth]{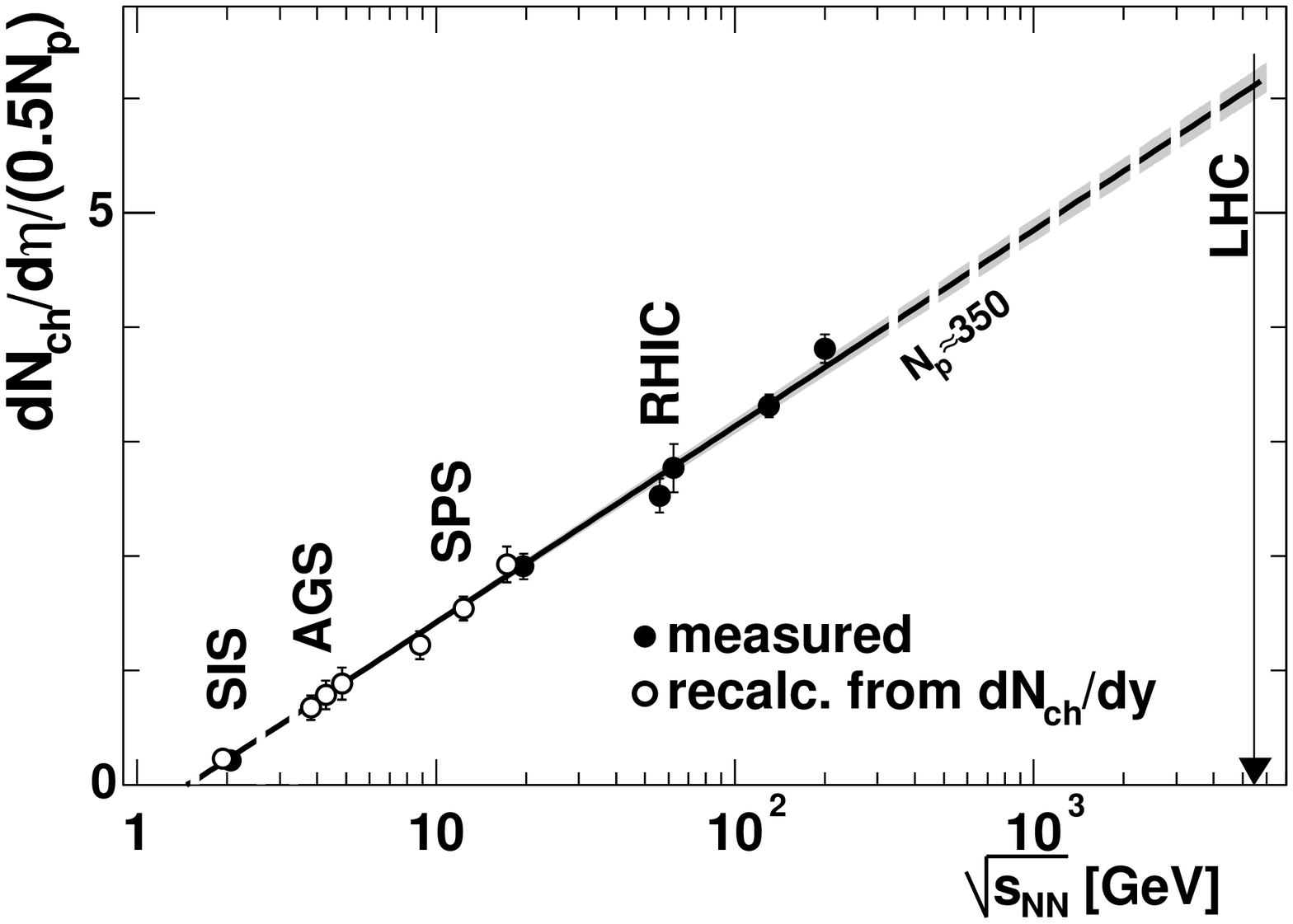}
\caption{Left panel: \dEt divided by the number of \Np pairs measured in the most 
central bin (value given in brackets) as a function of incident nucleon energy. The line is a logarithmic fit. 
The band corresponds to a 1$\sigma$ statistical deviation of the fit parameters. 
Right panel: the same for $dN_{ch}/d\eta$. The values of \Nch are the average values 
corresponding to \Nps~=~350. The single point at \sqn=~56~GeV is based on~\cite{phobos1}.\label{fig:sqn}}
\end{figure}

Figure~\ref{fig:sqn} shows the energy dependence for the \dEt and \dNch 
production per pair of participants in the most central collisions 
measured by different experiments. See~\cite{ppg19} for the details of the data compilation.

The results shown in Fig.~\ref{fig:sqn} are consistent with logarithmic 
scaling as described in~\cite{phenix_milov,sasha_thesis,david_thesis}. 
Use of the logarithmic function is phenomenological and is suggested 
by the trend of the data in the range of available measurements.
The agreement of the fits with the data in both panels is very good, especially in 
the right panel where the averaged values are used for \Np=350. The single point 
of NA49~\cite{na49_4} is excluded from the \Et fit. The results 
of the fit $dX/d\eta=(0.5N_{p}\cdot A) ln(\sqrt{s_{NN}}/\sqrt{s_{NN}^{0}})$ are:\\
for \Et  $\sqrt{s_{NN}^{0}} = 2.35\pm0.2$~GeV  and $A = 0.73\pm0.03$~GeV \\
for \Nch $\sqrt{s_{NN}^{0}} = 1.48\pm0.02$~GeV and $A = 0.74\pm0.01$.\\

The parameter $\sqrt{s_{NN}^{0}}$=2.35~GeV obtained from the \Et fit is 
slightly above, although within 3$\sigma$ from the minimum possible 
value of \sqns~=~$2\times a.m.u.$=1.86~GeV. The measurement closest to it 
at \sqns~=~2.05~GeV done by the FOPI experiment allows
to estimate the amount of \dEt produced to be 5.0~GeV in the most central collisions 
corresponding to \Np=359. Publication~\cite{ppg19} gives details of the estimate. 
This does not disagree with the extrapolation of the fit but 
does indicate that the logarithmic parameterization requires higher order 
terms to describe how the \Et production starts at very low \sqns.

The right panel of Fig.~\ref{fig:sqn} shows the logarithmic fit to the \Nch data. 
It agrees well with all \dNch results plotted for \Nps=350. Unlike that for $E_{T}$, 
the fit parameter $\sqrt{s_{NN}^{0}}$ for \Nch is 1.48$\pm$0.02~GeV which is 
lower than the minimum allowed \sqns. This suggests that above $2 \times a.m.u.$ 
the \Nch production as a function of \sqn should undergo threshold-like behavior, 
unlike the \Et production which must approach zero smoothly due to energy 
conservation. 

The FOPI measurement at \sqns=1.94~GeV and 2.05~GeV agrees with the 
extrapolation of the fit at energy very close to $2 \times a.m.u.$. It is an 
interesting result that colliding nuclei with kinetic energies of
0.037~GeV and 0.095~GeV per nucleon in the C.M.S. follow the same particle 
production trend as seen at AGS, SPS and RHIC energies.

A fit to the charged particle multiplicity shows a factor of 2.2 increase in \dNch 
per participant in the most central events from the highest energy at the AGS (\sqn=4.8~GeV) to the 
highest energy at the SPS (\sqn=\7) and a factor of 2.0 from the highest SPS energy to the highest 
RHIC energy (\sqn=\2). Assuming the same behavior extends to the LHC highest energy \sqns=5500~GeV one would expect  
\dNch = $(6.1 \pm 0.13)\cdot (0.5 N_{p})$ 
and the increase in particle production from the highest RHIC energy to be 
$\sim$60\% for the most central events. 
With the greater energy, the rapidity width should increase by $\sim$60\%
i.e. the total charged particle multiplicity would increase by 
a factor of $\sim$2.6 from the top RHIC energy.

It is interesting to compare the \sqn dependence shown in Fig.~\ref{fig:sqn} to the similar plot for 
$dN_{ch}/dy$ vs. \sqn published in~\cite{kink}. In Fig.4 of that publication the energy dependence 
of $dN_{ch}/dy$ for \Nps~=350 changes the slope between AGS 
and lower SPS energy. That is not seen in Fig.~\ref{fig:sqn}. The parameterization suggested in 
publication~\cite{kink}, $dN_{ch}/dy\propto(\sqrt{s_{NN}})^{0.3}$ is also different from the logarithmic 
scaling in Fig.~\ref{fig:sqn}. Although quantities in these two figures are not identical,
some contribution to the shape of the curve in Fig.4 in~\cite{kink} may be due to use of logarithmic 
scale on Y-axis, producing a bend-like shape out of straight line in a linear scale.

\begin{figure}[h]
\includegraphics[width=0.49\linewidth]{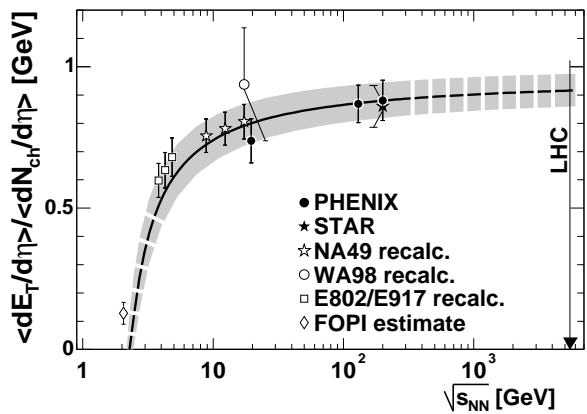}\hspace{2pc}%
\begin{minipage}[b]{18pc}
\caption{Ratio of \Et over \Nch for the most central events as a function 
of \sqn recalculated into C.M.S.. The line is the ratio of two fits shown 
in Fig.~\ref{fig:sqn}. The band corresponds to one standard deviation of 
the combined error.\label{fig:et_nc_sqn}}
\end{minipage}
\end{figure}

The ratio of \EN for the most central bin as a function of \sqn is 
shown in Fig.~\ref{fig:et_nc_sqn}.
Note that the line shown in the figure is not the fit to the data 
points. Rather, it is calculated as a ratio of the fits plotted in Fig.~\ref{fig:sqn}. 
The calculation agrees well with the data.

There are two regions in the plot which can be clearly separated. The
region from the lowest allowed \sqn to SPS energy is characterized
by a steep increase of the \EN ratio with \sqns. In this region the increase 
in the incident energy causes an increase in the \mt of the produced particles. 
The second region starts 
at the SPS energies. In this region, the \EN ratio is very weakly 
dependent on \sqns. The incident energy is converted into particle 
production at mid-rapidity rather than into increasing the particle $\langle m_{T} \rangle$. 

The shape of the \EN curve in the first region is governed by the 
difference in the $\sqrt{s_{NN}^{0}}$ parameter between \Et and $N_{ch}$. The 
second region is dominated by the ratio of the $A$ parameters in the fits; 
this ratio is close to 1~GeV. Extrapolating to LHC energies one gets
a \EN value of (0.92$\pm$0.06)~GeV.

The \EN parameter is usually associated with the chemical freeze-out temperature $T$. A simple relation 
like $2/3 E_{T}/N_{ch}\approx m_{0}+3/2 T$ finds $T\approx$0.22~GeV for a measured value \EN of 
0.88~GeV at top energy at RHIC. Coefficient 2/3 takes into account neutral particle contribution 
and the average particle mass $m_{0}\approx0.25$~GeV can be estimated from~\cite{tatsuja}. The freeze-out temperature estimate 
of $T\approx$0.22 GeV extracted this way is in an agreement to $T\approx$0.17 GeV measured from the 
spectra~\cite{tatsuja} plus some contribution of particle flow. In publications~\cite{Redlich,pbm} 
authors suggest the same \EN value close to 1~GeV is valid even at AGS and SIS energies. This is not in 
disagreement with the curve in Fig.~\ref{fig:et_nc_sqn} because the \Et definition used 
in~\cite{Redlich,pbm}, includes full baryon rest mass. We use different definition
of \Et as explained in~\cite{ppg19} where mass of pre-existing baryons is not included.
At lower energy where such contribution to the \Et is larger than at RHIC it would bring the \EN ratio closer to one.

The transition between two region with different \EN behavior occurs in the lower SPS energies. There are 
other observables which also undergo transition between regimes in the same range of \sqns. For example, the 
$\langle K^{+}\rangle /\langle \pi^{+}\rangle$ ratio demonstrates a peak at around \sqns~$\approx$8~GeV 
as shown in Fig.4 in reference~\cite{na49_recent}. In the same region, the volume of the colliding system 
goes through a minimum as shown in Fig.1 in reference~\cite{ceres}. The asymmetric flow parameter $v_{2}$ 
lies in-plane at and above the highest SPS energy and out-of-plane below it, see Fig.8 in 
publication~\cite{harry}. These evidences suggest that the energy range of 5--10~GeV has interesting 
physics potential which needs to be studied.

\subsubsection{Centrality shape}
\label{sec:centrality_shape}
Another interesting question is how the shapes of the centrality curves 
of \Et and \Nch change with \sqns. 

One approach, previously used in a number of papers is to describe the shape of the 
centrality dependence as a sum of ``soft'' and ``hard'' contributions such that the 
``soft'' component is proportional to \Np and the ``hard'' component to the the number of 
binary collisions $N_{c}$: $A \times N_{p} + B \times N_{c}$. A disadvantage of this 
approach is that the contributions called ``soft'' and ``hard'' do not necessarily 
correspond to the physical processes associated with these notations. 
Another approach is to assume that the production of \Et and \Nch is proportional to 
$N_{p}^{\alpha}$; although the parameter $\alpha$ does not have any physical meaning.

We present the results of $B/A$ and $\alpha$ obtained from the fits 
to the data at different \sqn in Table~\ref{tab:alphas}. Although the numbers 
tend to increase with beam energy, the values presented are consistent with each other 
within the systematic errors.

\begin{center}
\begin{table}
\caption{``B/A'' ratio and parameter $\alpha$ from the fit to the data. Errors are 
calculated assuming a change of the slope of centrality curves within the limits of the ``tilt'' 
errors for PHENIX and full errors for the averaged data (Table~\ref{tab:averages}).\label{tab:alphas}}
\centering
\begin{tabular}{cccc}
\br
\sqn &        \dEt         &           \dNch       &      \dNch         \\
 GeV &        PHENIX       &          PHENIX       &      Average       \\
\multicolumn{4}{c}{$B/A$}\\
\mr
200  & $0.49^{+.69}_{-.22}$ & $0.41^{+.57}_{-.21}$ &$0.28^{+.18}_{-.15}$\\
130  & $0.41^{+.52}_{-.23}$ & $0.41^{+.45}_{-.23}$ &$0.26^{+.18}_{-.11}$\\
19.6 & $0.37^{+.48}_{-.22}$ & $0.21^{+.30}_{-.15}$ &$0.23^{+.73}_{-.23}$\\
17.2 &			    &                      &$0.31^{+.46}_{-.24}$\\
8.7  &			    &                      &$0.12^{+.64}_{-.20}$\\
\multicolumn{4}{c}{parameter $\alpha$}\\
\hline
200  & 1.20$\pm$0.07 & 1.18$\pm$0.08 & 1.16$\pm$0.06\\
130  & 1.14$\pm$0.08 & 1.17$\pm$0.08 & 1.14$\pm$0.05\\
19.6 & 1.13$\pm$0.07 & 1.09$\pm$0.06 & 1.10$\pm$0.11\\
17.2 &               &               & 1.11$\pm$0.08\\
8.7  &               &               & 1.06$\pm$0.13\\
4.8  &               &               & 1.20$\pm$0.24\\
\br
\end{tabular}
\end{table}
\end{center}

Higher quality data would make it possible to derive a more conclusive 
statement about the shape of the curves plotted in Figs.~\ref{fig:comp_rhic}~and~\ref{fig:sps_ags}. 
With the present set of data, usually limited to \Np above 50, 
a large part of the centrality curve is missing or smeared by systematic errors.
To avoid this, we compare $Au+Au$ collisions to $p+p$ (\Nps=2) at the same energy.

\begin{figure}[h]
\includegraphics[width=0.49\linewidth]{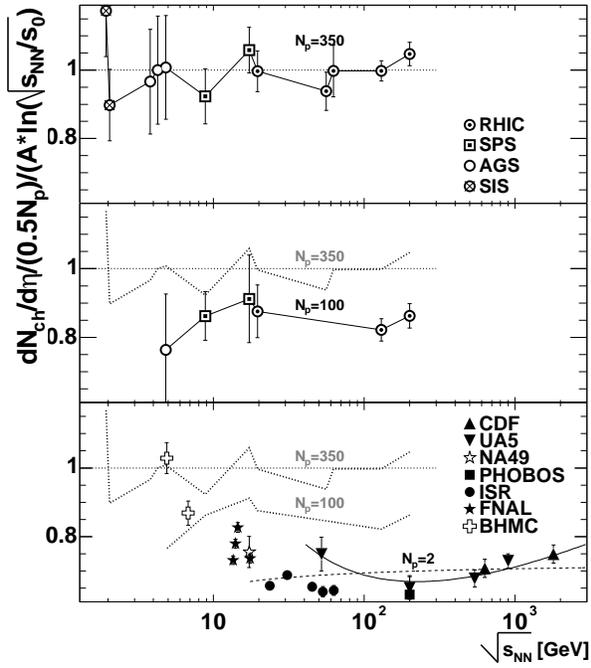}\hspace{2pc}%
\begin{minipage}[b]{18pc}
\caption{The three panels show $dN_{ch}/d\eta/(.5N_{p})$ divided by the logarithmic parameterization from 
Fig.~\ref{fig:sqn}. The panels correspond to \Nps~=~350,~100~and~2 ($p+p$) from top to bottom. 
$Au+Au$ points are connected with lines also shown in lower panels for comparison. 
The $Au+Au$ data is tabulated in Table~\ref{tab:averages}; $p+p$ data and parameterizations 
$dN/d\eta=2.5-0.25ln(s_{NN})+0.023ln(s_{NN})^{2}$ (solid line) and 
$dN/d\eta=0.27ln(s_{NN})-0.32$ (dashed line) are taken from~\cite{ua5,isr}. 
\label{fig:centr}}
\end{minipage}
\end{figure}

Figure~\ref{fig:centr} shows $dN_{ch}/d\eta/(0.5N_{p})$  
divided by the parameterization plotted in the right panel of Fig.~\ref{fig:sqn}.
The top panel shows the most central events with $N_{p}\approx$350. All points are consistent 
with one demonstrating an agreement of the fit to the data. The points are connected with 
a line for visibility. The middle panel shows results for mid-central events with \Nps~=~100 
connected with a solid line. The dashed line is the line from the top panel for \Nps~=~350.
The points for \Nps~=~100 are lower than \Nps~=~350 by a factor of $0.8-0.9$, over the plotted 
range of incident energies. The lower panel shows $p+p$ data corresponding to \Nps~=~2 measured 
by several experiments. Dashed lines are the same as appear in the upper two panels for \Nps~=~350~and~100 
and the $p+p$ parameterizations from~\cite{ua5,isr}. In the range of RHIC energies these 
points are lower by a factor of $0.65-0.75$ than the most central. Dotted lines show data from 
the upper two panels. 

These results indicate that the centrality curves normalized to the most central collisions 
have a similar shape for all RHIC energies within the errors of available measurements.

At lower energies the $p+p$ data show rise and become equal to the most central results. 
That would mean that at the lower SPS 
and AGS energies the centrality profile is independent of \Nps. At the same time the 
cross over takes place at AGS energy and it would suggest that below it the particle production 
per participant decreases with \Nps. There is no measurement supporting such a statement. Also, the 
the information about $p+p$ particle production at low energies is recalculated from total number 
of produced particles or identified particles sets which require additional assumptions and 
systematic uncertainties to be assigned to them. Hence, it is not as reliable as at higher energies.

 The data
for of the FNAL and BHMC collaborations are shown in Fig.~\ref{fig:centr} as an indication of the 
trend.

\subsection{Comparison to models}
A variety of models attempting to describe the behavior of \Et and \Nch as a function 
of centrality at different \sqn are available. 
An updated set of model results were collected from several theoretical groups 
to make a comparison as comprehensive as possible\footnote{For the JAM generator 
authors extracted data by running the code and assigning \Np from the Glauber model simulations.}.

\begin{figure}[h]
\includegraphics[width=0.99\linewidth]{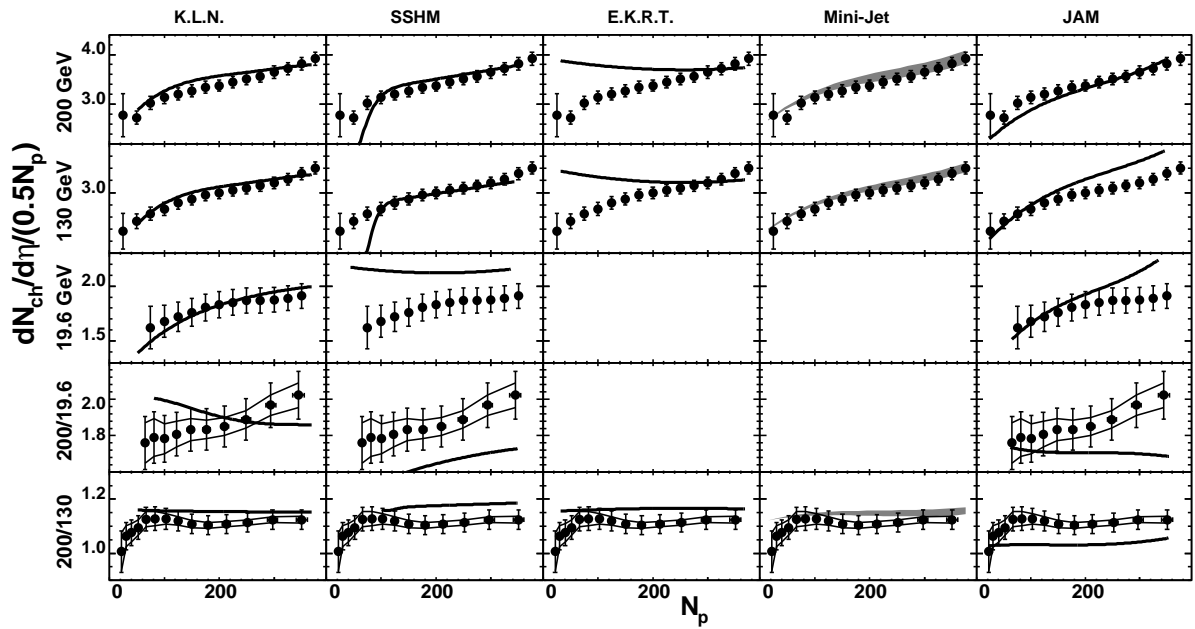}
\caption{\dNch per pair of participants compared to theoretical models. 
{\it KLN}~\cite{kln}, {\it SSHM}~\cite{accardi}, {\it EKRT}~\cite{ekrt}, 
{\it Minijet}~\cite{minijet} and {\it JAM}~\cite{jam}. The band shows 
the range of prediction for the Minijet model.\label{fig:theo1}}
\end{figure}

\begin{figure}[h]
\includegraphics[width=0.99\linewidth]{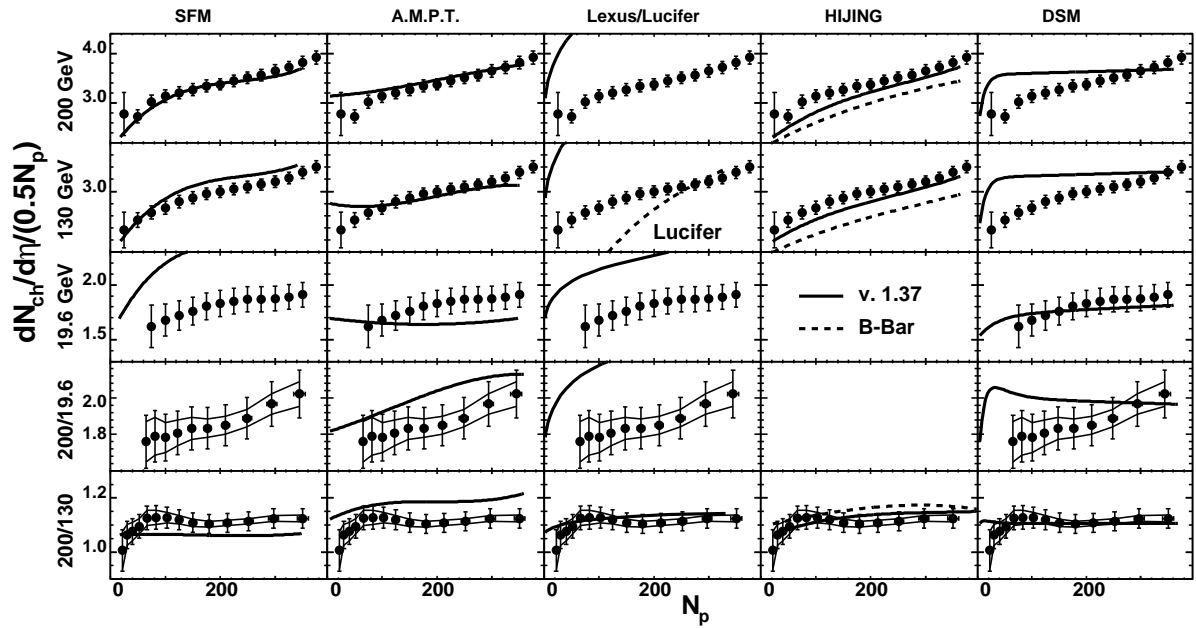}
\caption{Theoretical models compared to \dNch per pair of participants. {\it SFM}~\cite{perez}, {\it AMPT}~\cite{ampt}, {\it LEXUS}~\cite{lexus},{\it LUCIFER}~\cite{kahana},{\it HIJING}~\cite{hijing_sim, new_hijing} and {\it DSM}~\cite{dsm}. \label{fig:theo2}}
\end{figure}

\begin{figure}[h]
\includegraphics[width=0.99\linewidth]{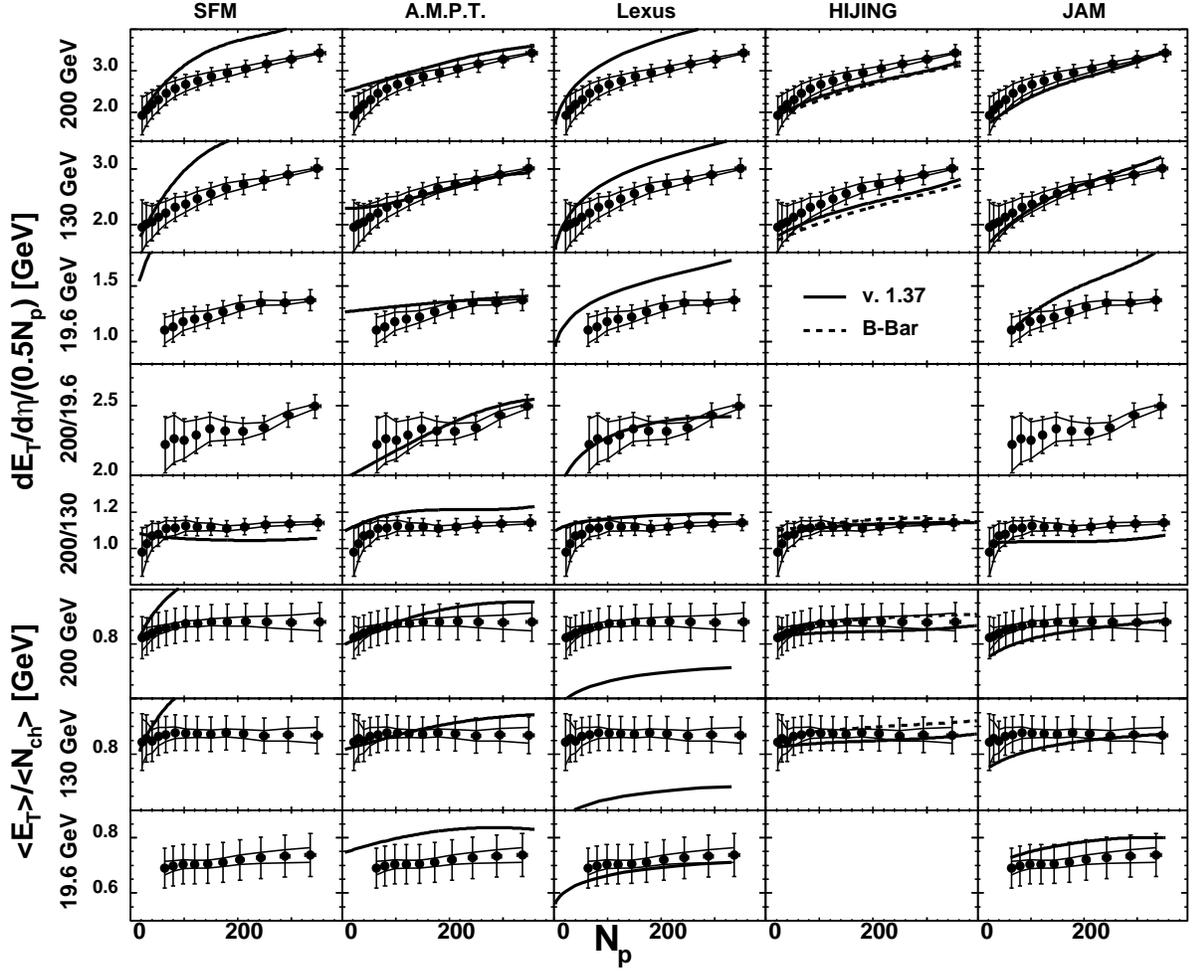}
\caption{Theoretical models compared to \dEt per pair of participants (upper panels) and per produced charged particle (lower panels). {\it SFM}~\cite{perez}, {\it AMPT}~\cite{ampt}, {\it LEXUS}~\cite{lexus} and {\it HIJING}~\cite{hijing_sim, new_hijing}.\label{fig:theo_ratio1}}
\end{figure}

Figures~\ref{fig:theo1} 
through~\ref{fig:theo_ratio1} show the comparison between the existing theoretical 
models\footnote{Models 
are presented as the best fit by the polynomial of the lowest degree which 
is closer than 1\% to any theoretical point. 
The polynomial is plotted in the range where points are provided.}
and the data for 19.6, 130 and 200 GeV. Brief descriptions of the models and 
their main characteristics are given below.

One of the more commonly used Monte Carlo event generators 
is {\it HIJING}~\cite{hijing_sim, new_hijing}. This model, like several others,
uses pQCD for initial minijet production, and the Lund string 
model~\cite{lund_string} for jet fragmentation and hadronization. 
{\it HIJING} also includes jet quenching and 
nuclear shadowing. 
This type of model typically has two components, a soft part  
proportional to \Np and a hard part proportional to $N_{coll}$, which
partly motivated the discussion in section~\ref{sec:centrality_shape}.  
There are also so-called 
saturation models which also rely on pQCD and predict that at some fixed scale
the gluon and quark phase-space density saturates, thus limiting the number of
produced quarks and gluons. An example of this type of model is 
{\it EKRT}~\cite{ekrt}, which is referred to as a final state saturation model.
In this paper, comparisons are also made to another parton 
saturation type model, {\it KLN}~\cite{kln}, an
initial state saturation model, and also to models related to {\it HIJING}, 
namely 
{\it Minijet}~\cite{minijet} and {\it AMPT}~\cite{ampt}. {\it AMPT} is a 
multiphase transport model, and extends {\it HIJING} by including explicit 
interactions between initial minijet partons and also final state hadronic interactions. 
{\it Minijet} follows the same two-component model
as {\it HIJING} but also incorporates an energy dependent cut-off scale,
similar to the saturation models.
{\it JAM}~\cite{jam} uses {\it RQMD} and {\it UrQMD} inspired ideas for the low energy 
interactions and above resonance region soft string excitation is implemented along the lines of {\it HIJING}.

The other models are listed briefly below. 
{\it SSHM} and {\it SFM} did not have a designated short identifier, so
 they were named somewhat arbitrarily here, based on the physics the models 
incorporate.
{\it SSHM (Saturation for Semi-Hard Minijet)}~\cite{accardi} is also a 
two-component model: pQCD-based for semi-hard partonic interactions, while 
for the soft particle production it uses the wounded nucleon model.
{\it DSM}~\cite{dsm}, the Dual String Model, is basically the Dual 
Parton Model~\cite{dpm}, with the inclusion of strings. 
{\it SFM (String Fusion Model)}~\cite{perez}, is a string model which 
includes hard collisions, collectivity in the initial state (string fusion), 
and rescattering of the produced secondaries. 
Finally, there are the hadronic models, {\it LUCIFER}~\cite{kahana}, a cascade model, with input fixed
 from lower energy data, and 
{\it LEXUS}~\cite{lexus}, a Linear EXtrapolation of Ultrarelativistic 
nucleon-nucleon Scattering data to nucleus-nucleus collisions.

The available model results range from predicting (or postdicting) \dNch 
at one energy to predicting both
\dNch and \dEt at 19.6, 130 and 200 GeV. The models have varying success in 
reproducing the data. 

Figure~\ref{fig:theo1} shows that {\it KLN} model is one of the most successful at 
describing the \dNch centrality dependence for all three energies. 
However, at \sqn=\1 the theoretical curve is steeper than the data. 
This results in a reversed 
centrality dependence relative to the data for the \2 to \1 ratio.
{\it SSHM} describes the 130 and 200 GeV data well for centralities 
above \Np$\sim$100, which is the approximate limit of applicability for 
this and other 
saturation models. For the less central events, the models are lower 
than the data. At 19.6 GeV, the model values are 
significantly higher than the data. 
The saturation model
{\it EKRT} describes the central points at both energies but overshoots the 
more peripheral data points and thus does not reproduce the general centrality 
dependence of the data. 
For the non-saturation models included in this figure, {\it Minijet} reproduces both the 
overall scale, as well as the centrality and energy dependence of 
the data rather well.

The {\it JAM} model shown in Fig.~\ref{fig:theo1} and most of the models included in Fig.~\ref{fig:theo2},
provided values for all three energies: 19.6, 130 and 200 GeV. 
{\it JAM} is rather successful in describing centrality shape at \sqns~=200~GeV. 
It also consistent with the measurements at \3 and partially at 19.6~GeV. In ratios it underestimates the data. 
For the \Et results shown in Fig.~\ref{fig:theo_ratio1} {\it JAM} reproduced the centrality shapes well, 
but underestimates the \sqn dependence of the data.
{\it SFM} is in reasonable agreement with the 130 and 200 GeV data, but gives
much larger values than the data at 19.6 GeV.
{\it AMPT} is overall in good agreement with the data for the two higher energies,
except for the increasing trend in \dNch at the most peripheral events, which
is not seen in the data. At the lower energy the \Nch centrality behavior is 
underestimated. 
{\it LEXUS} severely 
overshoots the data for all energies, indicating
that nucleus-nucleus effects are not correctly accounted for.
{\it LUCIFER} describes the central 
points at 130 GeV well, but undershoots the less central values at this energy.
The {\it HIJING} models (version 1.37 and a new version with implemented baryon junctions, {\it HIJING B-$\bar{B}$}) 
only provide points at 130 and 200 GeV and are in 
reasonable agreement with the data at those energies, but generally give 
somewhat lower values. The curves shown include quenching and shadowing {\it HIJING}.
{\it DSM} describes 19.6 GeV reasonably well for all centralities, and the more
central bins for 130 and 200 GeV, but overpredicts the values for 
semi-central and peripheral events.

Figure~\ref{fig:theo_ratio1} shows the results for the models that provide 
data for both \dNch and $dE_{T}/d\eta$. For $dE_{T}/d\eta$, {\it LEXUS} and {\it SFM} consistently overshoot the 
data for all energies. In the ratio \EN, {\it LEXUS} gives values that are 
too low except at the lowest energy 19.6 GeV. That indicates that the 
hadronization mechanism allows too little energy per particle. 
The {\it SFM} gives values that are too large, except for the most peripheral bin, 
this suggests, that the particles are assigned transverse masses that are too large.
The {\it HIJING} versions and the related {\it AMPT} model are in reasonable 
agreement with the data for both \dEt and $E_{T}/N_{ch}$\footnote{Note that the 
{\it HIJING} versions available at the time the data were collected and used
for predictions were in worse agreement with the data~\cite{phenix_bazik}. This was
before energy loss and minijet separation/cut-off scale parameters were updated.}.

Also shown in Fig.~\ref{fig:theo_ratio1} are the ratios of results at 
\2 to \1, and 
\2 to \3, for $dE_{T}/d\eta$. These results, especially the comparison of the \2 to \1 data, is intended 
to make a 
more precise check of the \sqn dependence of the models. 
{\it SFM} fails to describe the \1 data and thus can not describe
the energy dependence
probed by these ratios, unlike {\it LEXUS} which however does not agree
well with the individual data curves for 19.6, 130 and 200 GeV.
{\it AMPT} and the {\it HIJING} versions reproduce the values of the ratios well,
as expected since they are in reasonable agreement with the individual curves.
{\it AMPT} and {\it HIJING} are also successful in describing the \EN ratio, as illustrated
in the lower panels of Fig.~\ref{fig:theo_ratio1}.

To summarize, most models reproduce at least some of the data fairly well,
but most also fail in describing all the data. 
Since the model results typically are given without systematic errors, it is 
not entirely straightforward to quantify the level of agreement or 
disagreement with the data. Qualitatively, the models that are most successful 
in describing both \dEt and \dNch in terms of the overall trends, both 
regarding centrality dependence and energy dependence, are {\it AMPT}, 
and the {\it HIJING} versions. {\it KLN} and {\it Minijet} unfortunately do not give 
information on \dEt but are successful in describing the \dNch results.
The \dNch results thus can either be described by the initial state saturation scenario ({\it KLN}) or
by the mini-jet models that need an energy-dependent mini-jet   
cut-off scale as described in~\cite{new_hijing, minijet} to reproduce the data.

\section{Summary\label{sec:summary}}

This paper presents a systematic study of the energy and centrality
dependence of the charged particle multiplicity and transverse energy 
at mid-rapidity at \sqn = 19.6, 130 and 200 GeV. 

The yields, divided by the number of participant nucleons, show a consistent
centrality dependence (increase from peripheral to central) between \dEt and 
\dNch for all energies. Furthermore, the increase in the ratio \EN from \1 to \2 
is consistent with a 20\% increase in \mt with increasing \sqns. 
The ratio \EN shows only a weak centrality dependence at RHIC energies.

For the \sqn dependence, comparisons were made not only among RHIC results
but also including data from lower energy fixed-target experiments at SPS, 
AGS and SIS. A phenomenological fit, scaling logarithmically with \sqns, 
describes both \dEt and \dNch well for all energies and for the most central collisions.

Using the fit results, one can separate two regions with different 
particle production mechanisms. The region below SPS energy is characterized by a 
steep increase in \EN$\sim$\mt with \sqns, whereas for the energies above 
SPS \EN is found to be weakly dependent on \sqns.

Within the systematic errors of the measurements 
the shape of the centrality curves of $dN_{ch}/d\eta/(0.5N_{p})$ vs. \Np 
were found to be the same in the range of RHIC energies and to scale with 
$\ln(\sqrt{s_{NN}})$. The same trend must be true for \Et because \EN has a very 
weak centrality dependence.

Based on the \dEt measurements, the Bjorken energy density 
estimates were performed and $\epsilon_{Bj} \cdot \tau$ was determined to be
$5.4 \pm 0.6$ GeV/($fm^2 \cdot c$) at \sqn = \2 for the most central bin. 
This is in excess of what is
believed to be sufficient for a phase transition to the new state of matter. 
The energy density increases by about a factor of 2.6 from SPS to RHIC energy.

Finally, a comparison between the RHIC \dNch and \dEt data and a collection 
of models was performed. A few models, notably {\it HIJING} and {\it AMPT},
reproduce both \dEt and \dNch rather well for several energies.



\section*{References}
\begin{thebibliography}{99}

\bibitem{ppg19} PHENIX Collaboration S.S.~Adler \etaL, 
  nucl-ex/0409015

\bibitem{rhic} H.~Hahn \etaL, 
  \Journal{\NIMA} {499}{245-263}{2003}.

\bibitem{phenix} PHENIX Collaboration, K.~Adcox \etaL, 
  \Journal{\NIMA} {499}{469-479}{2003}.

\bibitem{pc} K.~Adcox \etaL, 
  \Journal{\NIMA} {497}{263-293}{2003}.

\bibitem{emcal} L.~Aphechetche \etaL, 
  \Journal{\NIMA} {499}{521-536}{2003}.

\bibitem{bbc-zdc} M.~Allen \etaL, 
  \Journal{\NIMA} {499}{549-559}{2003}.

\bibitem{zdc} S.~Adler \etaL, 
  \Journal{\NIMA} {470}{488-499}{2001}.

\bibitem{phenix_et} PHENIX Collaboration, K.~Adcox \etaL,
 	\Journal{\PRL} {87}{052301}{2001}.

\bibitem{phenix_nch} PHENIX Collaboration, K.~Adcox \etaL,
   	\Journal{\PRL} {86}{3500}{2001}.

\bibitem{phenix_milov} A.~Milov for the PHENIX Collaboration, 
	\Journal{\NPA} {698}{171}{2002}.

\bibitem{phenix_bazik}  A.~Bazilevsky for the PHENIX Collaboration,
	\Journal{\NPA} {715}{486}{2003}.

\bibitem{david_thesis} D.~Silvermyr, 
  Ph.D. Thesis. Lund University (2001).

\bibitem{sasha_thesis} A.~Milov, 
  Ph.D. Thesis. The Weizmann Institute of Science (2002).

\bibitem{julia} PHENIX Collaboration, K.~Adcox \etaL,
   	\Journal{\PRL} {88}{242301}{2002}.

\bibitem{geant} GEANT 3.2.1, 
  CERN program library.

\bibitem{hijing_sim} X.N.~Wang and M.~Gyulassy, 
 	\Journal{\PRD} {44}{3501}{1991}.

\bibitem{tatsuja} PHENIX Collaboration, S.S.~Adler \etaL, 
  \Journal{\PRC} {69}{034909}{2004}.

\bibitem{na49_1} J.~B\"{a}chler for the NA49 collaboration,
	\Journal{\NPA} {661}{45}{1999}.

\bibitem{na49_2}  M.~van Leeuwen for the NA49 collaboration,
        \Journal{\NPA} {715}{161}{2003}.

\bibitem{na49_3}  NA49 Collaboration, S.V.~Afanasiev \etaL, 
        \Journal{\PRC} {66}{054902}{2002}. 

\bibitem{PDG} K.~Hagiwara \etaL, http://pdg.lbl.gov,
	\Journal{\PRD} {66}{010001}{2002}.

\bibitem{Hofstadter} B.~Hahn, D.G.~Ravenhall and R.~Hofstadter,
	Phys. Rev, 101, 1131 (1956), and\\
	C.W.~De Jager \etaL, Atomic Data and Nuclear Table 24, 479 (1974).

\bibitem{phobos6} PHOBOS Collaboration, B.B.~Back \etaL, 
        nucl-ex/0210015.

\bibitem{ags_mjt} E-802 Collaboration, T.~Abbott \etaL,
        \Journal{\PRC} {63}{064602}{2001}.

\bibitem{mitchell} PHENIX Collaboration, K.~Adcox \etaL,
        \Journal{\PRC} {66}{024901}{2002}.

\bibitem{wa98} WA98 Collaboration, M.M. Aggarval \etaL, 
	\Journal{\EPJC} {18}{651}{2001}.

\bibitem{phenix_pp} PHENIX Collaboration, S.S.~Adler \etaL,
        \Journal{\PRL} {91}{241803-1}{2003}.


\bibitem{gulash} M.~Gyulassy, 
  nucl-th/0106072.

\bibitem{raju} A.~Krasnitz, Y.~Nara, R.~Venugopalan, hep-ph/0305112,
   	\Journal{\NPA} {727}{427-436}{2003}.

\bibitem{bjorken} J.D.~Bjorken, 
        \Journal{\PRD} {27}{140}{1983}.

\bibitem{na49_4}  NA49 Collaboration, T.~Alber \etaL, 
        \Journal{\PRL} {75}{3814}{1995}.

\bibitem{kink} A.Andronic and P.Braun-Munzinger, hep-ph/04022091

\bibitem{Redlich} J.~Cleymans and K.~Redlich,
        \Journal{\PRL} {81}{5284}{1988}.

\bibitem{pbm} P.Braun-Munzinger, K.Redlich and J.Stachel 
  nucl-th/0304013

\bibitem{na49_recent} M.Gazdzicki for the NA49 Collaboration
  nucl-ex/0403023

\bibitem{ceres} CERES Collaboation D.Adamova \etaL,
        \Journal{\PRL} {90}{022301}{2003}.

\bibitem{harry} A Appelsh\"{a}user for the CERES Collaboation,
        \Journal{\NPA} {698}{253c-260c}{2002}.

\bibitem{star_bj} STAR Collaboration, J.~Adams \etaL, 
  nucl-ex/0311017.

\bibitem{brahms1} BRAHMS Collaboration, I.G.~Bearden \etaL, 
   	\Journal{\PLB} {523}{227}{2001}.

\bibitem{brahms2} BRAHMS Collaboration, I.G.~Bearden \etaL, 
 	\Journal{\PRL} {88}{202301}{2002}. 

\bibitem{phobos1} PHOBOS Collaboration, B.B.~Back \etaL, 
 	\Journal{\PRL} {85}{3100}{2000}.

\bibitem{phobos2} PHOBOS Collaboration, B.B.~Back \etaL, 
 	\Journal{\PRC} {65}{061901}{2002}.

\bibitem{phobos3} M.D.~Baker \etaL for the PHOBOS Collaboration,, 
        \Journal{\NPA} {715}{65c-74c}{2003}.

\bibitem{star1}   T.S.~Ullrich for the STAR Collaboration, 
         nucl-ex/0305018.

\bibitem{star2}   T.S.~Ullrich for the STAR Collaboration, 
        \Journal{\NPA} {715}{399}{2003}.

\bibitem{star_etra} STAR Collaboration, J~ Adams \etaL, 
        nucl-ex/0407003.

\bibitem{ceres_3} D.~Mi\`{s}kowiec for the CERES Collaboration,
Proc. of CIPPQG Palaiseau, September 2001.

\bibitem{na50_1}  NA50 Collaboration, M.C.~Abreu \etaL, 
        \Journal{\PLB} {530}{43}{2002}.

\bibitem{na57}    NA57 Collaboration, F.~Antinori \etaL, 
        nucl-ex/0406004.


\bibitem{phobos_total} PHOBOS Collaboration, B.B.~Back \etaL, 
        nucl-ex/0301017


\bibitem{ua5}  UA5 Collaboration, G.J.~Alner \etaL,
	\Journal{\ZPC} {33}{1-6}{1986}.

\bibitem{isr}  W.~Thome \etaL, 
 	\Journal{\NPB} {129}{365-389}{1977}. 

\bibitem{fnal}  J.~Whitmore,
        \Journal{\PLC} {10}{274-373}{1974}.

\bibitem{bhdm} V.~Blobel \etaL,
 	\Journal{\NPB} {69}{454-492}{1974}. 

\bibitem{new_hijing} V.~Topor~Pop, M.~Gyulassy, J.~Barrette, C.~Gale,
       X.~N.~Wang, N.~Xu, and K.~Filimonov,
 	\Journal{\PRC} {68}{054902}{2003}.

\bibitem{lund_string} B.~Andersson \etaL,  
 	\Journal{\NPB} {281}{289}{2003}.

\bibitem{ekrt} K.J.~Eskola \etaL, 
	\Journal{\NPB} {570}{379}{2000};
   	\Journal{\PLB} {497}{39}{2001}.

\bibitem{kln} D.~Kharzeev and M.~Nardi, 
   	\Journal{\PLB} {507}{121}{2001};
	D.~Kharzeev and E.~Levin, 
   	\Journal{\PLB} {523}{79}{2001}.
 
\bibitem{minijet} S.~Lee and X.N.~Wang,
   	\Journal{\PLB} {527}{85}{2002}. 

\bibitem{jam} Y.Nara,
   	\Journal{\NPA} {638}{555c}{1998}. 

\bibitem{ampt} Z.~Lin \etaL,
   	\Journal{\PRC} {64}{011902}{2001}. 

\bibitem{accardi} A.~Accardi,
   	\Journal{\PRC} {64}{064905}{2001}. 

\bibitem{dsm} R.~Ugoccioni,
   	\Journal{\PLB} {491}{253}{2000}. 

\bibitem{dpm} A.~Capella {\it et al},
   	\Journal{\PLC} {236}{225}{1994}. 

\bibitem{perez} N.~Armesto Perez \etaL,
   	\Journal{\PLB} {527}{92}{2002}; 
   	\Journal{\EPJC} {22}{149}{2001}. 

\bibitem{kahana} D.E.~Kahana and S.H.~Kahana,
  nucl-th/0208063.

\bibitem{lexus} S.~Jeon and J.~Kapusta,
   	\Journal{\PRC} {63}{011901}{2001}. 


\bibitem{ceres_1} F.~Ceretto,  
  Ph.D. Thesis. University of Heidelberg (1998).

\bibitem{ceres_2} H.~Appelsh\"{a}user for the CERES Collaboration,
        \Journal{\NPA} {698}{253}{2002}.

\bibitem{e802_1} E802 Collaboration, L.~Ahle \etaL, 
 	\Journal{\PRC} {59}{2173}{1999}. 

\bibitem{e802_2} E802 Collaboration, L.~Ahle \etaL, 
 	\Journal{\PRC} {58}{3523-3538}{1998}. 

\bibitem{e802_3} C.A.~Ogilvie for the E866 and E917 Collaborations,
 	\Journal{\NPA} {638}{57c-68c}{1998}. 

\bibitem{e802_5} Y.~Akiba for the E802 Collaboration,
        \Journal{\NPA} {610}{139}{1996}.

\bibitem{e917_4} E917 Collaboration, B.B.~Back \etaL, 
 	\Journal{\PRL} {86}{1970-1973}{2001}. 

\bibitem{e802_6} E802 Collaboration, L.~Ahle \etaL, 
 	\Journal{\PRC} {57}{R466-R470}{1998}. 

\bibitem{e802_7} E802 Collaboration, L.~Ahle \etaL, 
 	\Journal{\PLB} {476}{1}{2000}. 

\bibitem{e802_8} E802 Collaboration, L.~Ahle \etaL, 
 	\Journal{\PLB} {490}{53}{2000}. 

\bibitem{fopi_1} FOPI Collaboration, W.~Reisdorf \etaL, 
 	\Journal{\NPA} {612}{493}{1997}. 

\bibitem{fopi_2} FOPI Collaboration, D.~Pelte \etaL, 
 	\Journal{\ZPA} {357}{215}{1997}. 

\bibitem{fopi_3} FOPI Collaboration, B.~Hong \etaL, 
 	\Journal{\PRC} {66}{034901}{2002}. 
 
\end{thebibliography}
\end{document}